\documentclass[twocolumn]{aastex631}
\usepackage{amsmath}
\usepackage{epsf}
\usepackage{color}
\usepackage{graphicx}
\usepackage{color}
\usepackage{enumitem}
\usepackage{mathtools}
\usepackage[mathcal]{eucal}
 \let\mathscr\relax
\usepackage[scr]{rsfso}

\shorttitle{CEERS Key Paper VII}

\newcommand{\fagn}{$f({\rm AGN})_{\rm MIR}$}

\begin{document}

\title{CEERS Key Paper VII: JWST/MIRI Reveals a Faint Population of Galaxies at Cosmic Noon Unseen by Spitzer}

\author[0000-0002-5537-8110]{Allison Kirkpatrick}
\affiliation{Department of Physics and Astronomy, University of Kansas, Lawrence, KS 66045, USA}

\author[0000-0001-8835-7722]{Guang Yang}
\affiliation{Kapteyn Astronomical Institute, University of Groningen, P.O. Box 800, 9700 AV Groningen, The Netherlands}
\affiliation{SRON Netherlands Institute for Space Research, Postbus 800, 9700 AV Groningen, The Netherlands}

\author[0000-0002-9466-2763]{Aur{\'e}lien Le Bail}
\affiliation{Universit{\'e} Paris-Saclay, Universit{\'e} Paris Cit{\'e}, CEA, CNRS, AIM, 91191, Gif-sur-Yvette, France}

\author{Greg Troiani}
\affiliation{Department of Physics and Astronomy, University of Kansas, Lawrence, KS 66045, USA}

\author[0000-0002-5564-9873]{Eric F.\ Bell}
\affiliation{Department of Astronomy, University of Michigan, 1085 S. University Ave, Ann Arbor, MI 48109-1107, USA}

\author[0000-0001-7151-009X]{Nikko J. Cleri}
\affiliation{Department of Physics and Astronomy, Texas A\&M University, College Station, TX, 77843-4242 USA}
\affiliation{George P.\ and Cynthia Woods Mitchell Institute for Fundamental Physics and Astronomy, Texas A\&M University, College Station, TX, 77843-4242 USA}

\author[0000-0002-7631-647X]{David Elbaz}
\affiliation{Universit{\'e} Paris-Saclay, Universit{\'e} Paris Cit{\'e}, CEA, CNRS, AIM, 91191, Gif-sur-Yvette, France}

\author[0000-0001-8519-1130]{Steven L. Finkelstein}
\affiliation{Department of Astronomy, The University of Texas at Austin, Austin, TX, USA}

\author[0000-0001-6145-5090]{Nimish P. Hathi}
\affiliation{Space Telescope Science Institute, Baltimore, MD, USA}

\author[0000-0002-3301-3321]{Michaela Hirschmann}
\affiliation{Institute of Physics, Laboratory of Galaxy Evolution, Ecole Polytechnique F\'ed\'erale de Lausanne (EPFL), Observatoire de Sauverny, 1290 Versoix, Switzerland}

\author[0000-0002-4884-6756]{Benne W. Holwerda}
\affil{Physics \& Astronomy Department, University of Louisville, 40292 KY, Louisville, USA}

\author[0000-0002-8360-3880]{Dale D. Kocevski}
\affiliation{Department of Physics and Astronomy, Colby College, Waterville, ME 04901, USA}

\author[0000-0003-1581-7825]{Ray A. Lucas}
\affiliation{Space Telescope Science Institute, 3700 San Martin Drive, Baltimore, MD 21218, USA}

\author[0000-0002-6149-8178]{Jed McKinney}
\affiliation{Department of Astronomy, The University of Texas at Austin, 2515 Speedway, Austin, TX, 78712, USA}

\author[0000-0001-7503-8482]{Casey Papovich}
\affiliation{Department of Physics and Astronomy, Texas A\&M University, College Station, TX, 77843-4242 USA}
\affiliation{George P.\ and Cynthia Woods Mitchell Institute for Fundamental Physics and Astronomy, Texas A\&M University, College Station, TX, 77843-4242 USA}

\author[0000-0003-4528-5639]{Pablo G. P\'erez-Gonz\'alez}
\affiliation{Centro de Astrobiolog\'{\i}a (CAB), CSIC-INTA, Ctra. de Ajalvir km 4, Torrej\'on de Ardoz, E-28850, Madrid, Spain}

\author[0000-0002-6219-5558]{Alexander de la Vega}
\affiliation{Department of Physics and Astronomy, University of California, 900 University Ave, Riverside, CA 92521, USA}

\author[0000-0002-9921-9218]{Micaela B. Bagley}
\affiliation{Department of Astronomy, The University of Texas at Austin, Austin, TX, USA}

\author[0000-0002-3331-9590]{Emanuele Daddi}
\affiliation{Universit{\'e} Paris-Saclay, Universit{\'e} Paris Cit{\'e}, CEA, CNRS, AIM, 91191, Gif-sur-Yvette, France}

\author[0000-0001-5414-5131]{Mark Dickinson}
\affiliation{NSF's National Optical-Infrared Astronomy Research Laboratory, 950 N. Cherry Ave., Tucson, AZ 85719, USA}

\author[0000-0001-7113-2738]{Henry C. Ferguson}
\affiliation{Space Telescope Science Institute, Baltimore, MD, USA}

\author[0000-0003-3820-2823]{Adriano Fontana}
\affiliation{INAF - Osservatorio Astronomico di Roma, via di Frascati 33, 00078 Monte Porzio Catone, Italy}

\author[0000-0002-5688-0663]{Andrea Grazian}
\affiliation{INAF--Osservatorio Astronomico di Padova, Vicolo dell'Osservatorio 5, I-35122, Padova, Italy}

\author[0000-0001-9440-8872]{Norman A. Grogin}
\affiliation{Space Telescope Science Institute, Baltimore, MD, USA}

\author[0000-0002-7959-8783]{Pablo Arrabal Haro}
\affiliation{NSF's National Optical-Infrared Astronomy Research Laboratory, 950 N. Cherry Ave., Tucson, AZ 85719, USA}

\author[0000-0001-9187-3605]{Jeyhan S. Kartaltepe}
\affiliation{Laboratory for Multiwavelength Astrophysics, School of Physics and Astronomy, Rochester Institute of Technology, 84 Lomb Memorial Drive, Rochester, NY 14623, USA}

\author[0000-0001-8152-3943]{Lisa J. Kewley}
\affiliation{Center for Astrophysics | Harvard \& Smithsonian, 60 Garden Street, Cambridge, MA 02138, USA}

\author[0000-0002-6610-2048]{Anton M. Koekemoer}
\affiliation{Space Telescope Science Institute, 3700 San Martin Dr., Baltimore, MD 21218, USA}

\author[0000-0003-3130-5643]{Jennifer M. Lotz}
\affiliation{Gemini Observatory/NSF's National Optical-Infrared Astronomy Research Laboratory, 950 N. Cherry Ave., Tucson, AZ 85719, USA}

\author[0000-0001-8940-6768]{Laura Pentericci}
\affiliation{INAF - Osservatorio Astronomico di Roma, via di Frascati 33, 00078 Monte Porzio Catone, Italy}

\author[0000-0003-3382-5941]{Nor Pirzkal}
\affiliation{ESA/AURA Space Telescope Science Institute}

\author[0000-0002-5269-6527]{Swara Ravindranath}
\affiliation{Space Telescope Science Institute, 3700 San Martin Drive, Baltimore, MD 21218, USA}

\author[0000-0002-6748-6821]{Rachel S. Somerville}
\affiliation{Center for Computational Astrophysics, Flatiron Institute, 162 5th Avenue, New York, NY, 10010, USA}

\author[0000-0002-1410-0470]{Jonathan R. Trump}
\affiliation{Department of Physics, 196 Auditorium Road, Unit 3046, University of Connecticut, Storrs, CT 06269, USA}

\author[0000-0003-3903-6935]{Stephen M.~Wilkins} %
\affiliation{Astronomy Centre, University of Sussex, Falmer, Brighton BN1 9QH, UK}
\affiliation{Institute of Space Sciences and Astronomy, University of Malta, Msida MSD 2080, Malta}

\author[0000-0003-3466-035X]{{L. Y. Aaron} {Yung}}
\altaffiliation{NASA Postdoctoral Fellow}
\affiliation{Astrophysics Science Division, NASA Goddard Space Flight Center, 8800 Greenbelt Rd, Greenbelt, MD 20771, USA}

\begin{abstract}
    The Cosmic Evolution Early Release Science (CEERS) program observed the Extended Groth Strip with the Mid-Infrared Instrument (MIRI) on the {\it James Webb Space Telescope} ({\it JWST}) in 2022. In this paper, we discuss the four MIRI pointings that observed with longer wavelength filters, including F770W, F1000W, F1280W, F1500W, F1800W, and F2100W. We compare the MIRI galaxies with the {\it Spitzer}/MIPS 24\,$\mu$m population in the EGS field. We find that 
    MIRI can observe an order of magnitude deeper than MIPS in significantly shorter integration times, attributable to {\it JWST}'s much larger aperture and MIRI's improved sensitivity. MIRI is exceptionally good at finding faint ($L_{\rm IR}<10^{10}\,L_\odot$) galaxies at $z\sim1-2$. 
     We find that a significant portion of MIRI galaxies are ``mid-IR weak''--they have strong near-IR emission and relatively weaker mid-IR emission, and most of the star formation is unobscured. We present new IR templates that capture how the mid-IR to near-IR emission changes with increasing infrared luminosity. We present two color-color diagrams to separate mid-IR weak galaxies and active galactic nuclei (AGN) from dusty star-forming galaxies and find that these color diagrams are most effective when used in conjunction with each other. 
     We present the first number counts of 10\,$\mu$m sources and find that there are $\lesssim10$ IR AGN per MIRI pointing, possibly due to the difficulty of distinguishing AGN from intrinsically mid-IR weak galaxies (due to low metallicities or low dust content). We conclude that MIRI is most effective at observing moderate luminosity ($L_{\rm IR}=10^9-10^{10}\,L_\odot$) galaxies at $z=1-2$, and that photometry alone is not effective at identifying AGN within this faint population. 
\end{abstract}

\section{Introduction}
In the past two decades, the Multi-Band Imaging 
Photometer (MIPS) on the {\it Spitzer Space Telescope} 
revealed a new understanding of the infrared Universe 
through observations of a previously undetected 
population of infrared-luminous galaxies at $z>0.5$ 
\citep{chary2004,papovich2004}. MIPS observations 
demonstrated that the bulk of star formation in the 
Universe occurred when the Universe was a mere $3-6$ 
billion years old ($z\sim1-2$), the so-called ``cosmic 
noon'' epoch \citep{perez2005,caputi2007}. Furthermore, 
contrary to the local Universe, this star formation 
occurred predominantly in dust rich galaxies--namely, 
(Ultra) Luminous Infrared Galaxies \citep[(U)LIRGs, $L_{\rm 
IR}>10^{11}\,L_\odot$, where $L_{\rm IR}$ is the integrated luminosity from $8-1000\,\mu$m;][]{magnelli2011}. Although 
such galaxies exist today, they contribute relatively 
little to the local star formation rate \citep{rodighiero2010}.

While (U)LIRGs at cosmic noon are forming stars at prodigious rates (star formation rates, SFRs, of $100-1000\,M_\odot/$yr), they predominately lie on the main sequence of star formation, in stark contrast to local (U)LIRGs \citep{elbaz2011,kartaltepe2012}. The main sequence is an empirical tight correlation between the SFR and stellar masses ($M_\ast$) of galaxies \citep{noeske2007,elbaz2007}. Galaxies on the main sequence are presumed to be undergoing long-lived secular evolution, while galaxies above the main sequence are undergoing short bursts of star formation, likely triggered by a major merger event. The Infrared Spectrograph (IRS) instrument on {\it Spitzer} was used to probe the mid-IR emission of these cosmic noon (U)LIRGs in detail and found that their mid-IR dust emission is remarkably similar to isolated LIRGs in the local Universe, indicating a metal-enriched environment and the distinct lack of an obscured nuclear starburst \citep{sajina2012,kirkpatrick2012,kirkpatrick2015,pope2013,mckinney2020}. Simply put, dusty star-forming galaxies in the distant Universe resemble scaled up versions of non-merging star-forming galaxies in the local Universe.

All massive galaxies are presumed to host a supermassive black hole at their centers. The peak of the supermassive black hole accretion rate density also occurs at cosmic noon \citep{delvecchio2014,peca2023}, making this epoch of fundamental importance to understanding the concurrent buildup of stellar and black hole mass. Furthermore, there is evidence from both deep X-ray and IR observations that the bulk of black hole growth at cosmic noon is obscured \citep{delmoro2016,ananna2019,yang2023}. Mid-IR emission can reliably be used to identify both obscured and unobscured active galactic nuclei \citep[AGN;][]{padovani2017,yang2023}. In previous mid-IR surveys with ${\it Spitzer}$/MIPS or the Wide-Field Survey Infrared Explorer (WISE), AGN make up at least 15\% of extragalactic sources with $S_{24}>100\mu$Jy \citep[hereafter K17]{jarrett2011,kirkpatrick2017}.

\begin{deluxetable*}{cllrcc}[ht!]
\tablecaption{CEERS Red MIRI Pointings}
\label{tbl}
\tablehead{\colhead{Pointing}\tablenotemark{a} & \colhead{RA} & \colhead{Dec} & \colhead{Filter} & \colhead{Exp. Time} & \colhead{$5\sigma$ depth}\tablenotemark{b} \\
\colhead{} & \colhead{(J2000)} & \colhead{(J2000)} & & \colhead{(sec)} & \colhead{($\mu$Jy)}}
\startdata
1 & 14:20:38.88 & +53:03:04.6 & 770W & 1648 & 0.21\\
 & & & 1000W & 1673 & 0.42 \\
 & & & 1280W & 1673 & 0.78 \\
 & & & 1500W & 1673 & 1.27\\
 & & & 1800W & 1698 & 2.51 \\
 & & & 2100W & 4812 & 4.79\\
2 & 14:20:17.42 & +52:59:16.2 & 770W & 1648 & 0.21 \\
 & & & 1000W & 1673 & 0.43 \\
 & & & 1280W & 1673 & 0.82 \\
 & & & 1500W & 1673 & 1.15 \\
 & & & 1800W & 1698 & 2.73 \\
 & & & 2100W & 4812 & 3.87 \\
5 & 14:19:03.74 & +52:49:08.5 &  1000W & 1243 & 0.50 \\
& & & 1280W & 932 & 1.27 \\
& & & 1500W & 932 & 2.31 \\
& & & 1800W & 1243 & 3.98\\
8 & 14:19:21.78 & +52:48:55.5 & 1000W  & 1243 & 0.47\\ 
& & & 1280W & 932 & 1.32\\
& & & 1500W & 932 & 2.00\\
& & & 1800W & 1243 & 3.98
\enddata
\tablenotetext{a}{For layout of the pointings, see https://ceers.github.io/obs.html}
\tablenotetext{b}{Flux limit calculations are derived from the AB magnitude limits presented in \citet{yang2023b}}.
\end{deluxetable*}

In star-forming galaxies, the mid-IR emission is dominated by polycyclic aromatic hydrocarbon (PAH) complexes at $6.2, 7.7, 11.2,$ and 12.7\,$\mu$m. AGN typically have a dusty toroidal structure surrounding the accretion disk that radiates as a power-law in the mid-IR. These differences make the mid-IR advantageous for identifying populations of star-forming galaxies and AGN, and several color-color methods exist for doing so \citep{lacy2004,stern2005,donley2012,kirkpatrick2013,messias2012}. The Mid-Infrared Instrument (MIRI) \citep{rieke2015} on the {\it James Webb Space Telescope} \citep[\textit{JWST};][]{gardner2006} has several improvements over MIPS: the greatly improved spatial resolution allows for dramatically improved source deblending, and the MIRI filter set better spans the mid-IR emission. These improvements led to studies predicting that MIRI would be efficient and reliable at finding AGN \citep{messias2012,langeroodi2023} at high redshift \citep[$z>5$;][]{volonteri2017} or at lower luminosities and high obscuration levels \citep{kirkpatrick2017,yang2021}.

In this paper, we explore the nature of the sources in one of the first MIRI surveys, which was observed as part of the Cosmic Evolution Early Release Science (CEERS; PI S.\,Finkelstein) program. In Section \ref{sec:data}, we discuss the observations. In Section \ref{sec:results}, we compare the MIRI sources to the MIPS sources in the Extended Groth Strip (EGS) field, and we discuss the presence of AGN. We discuss the implications of the MIRI observations, including the surprising lack of AGN, in Section \ref{sec:discussion}. Finally, we present our conclusions in Section \ref{sec:conclusions}.

In this work, we assume a standard flat cosmology with $H_0=70\,$Mpc/km/s, $\Omega_M=0.3$, and $\Lambda=0.7$.

\section{Data}
\label{sec:data}

\subsection{MIRI Observations}
The CEERS MIRI observations, data reduction, and photometric catalogs are fully described in \cite{yang2023}; here, we summarize relevant details. The EGS field was observed with eight MIRI pointings. Four ``blue'' pointings were only observed in the F560W and F770W filters. In this paper, we discuss the four ``red'' pointings, observed in the longer wavelength filters (Table \ref{tbl}). Pointings 1 (Figure 1) and 2 were observed in July 2022, while pointings 5 and 8 were observed in December 2022. Pointings 1 and 2 do not overlap with CEERS NIRCam observations. The F770W and F2100W filter are only observed in pointings 1 and 2. Table \ref{tbl} lists the 5$\sigma$ flux limits for each filter in each pointing \citep{yang2023b}. 
The MIRI FOV is $112.6''\times73.5''$, bringing the total observed area to 9.2\,arcmin$^2$ for the 4 MIRI pointings discussed in this work.

\begin{figure*}
    \includegraphics[width=\linewidth]{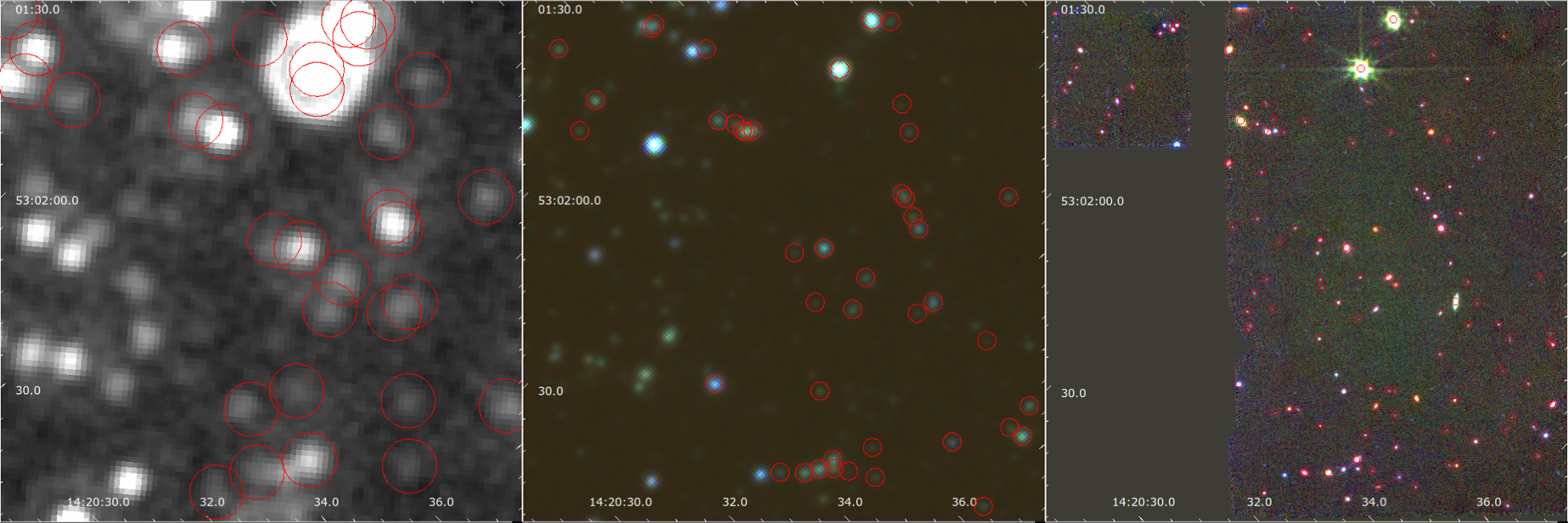}
    \caption{We show MIRI pointing 1 ({\it right panel}) alongside the {\it Spitzer}/IRAC ({\it middle}) and MIPS ({\it left}) observations of the same region. The apertures show the location of detected sources in each image (MIRI region only). For the MIPS (IRAC) image, the apertures are 6'' (2''), corresponding to the instrument beam size. 
    In the IRAC image, blue corresponds to channel 1 (3.6\,$\mu$m), green corresponds to channel 2 (4.5\,$\mu$m), and
red corresponds to channel 3 (5.8\,$\mu$m). In the MIRI image, the 770W filter is blue, F1000W is green, and F1280W is red.}
    \label{fig:image}
\end{figure*}

The MIRI photometry was
extracted with {\tt T-PHOT} \citep{merlin2015} using {\it Hubble
Space Telescope} ({\it HST})/F160W observations as a prior. HST positions were used rather than NIRCam due to the lack of uniform NIRCam coverage in the MIRI pointings. 
For details of the MIRI observations, data reduction, photometry extraction, and quality assessment see \citet{yang2021,papovich2023} and the CEERS MIRI observation paper \citep{yang2023b}.

The use of near-IR {\it HST} priors resulted in some spurious detections where a galaxy is visible in the {\it HST} image but not in all of the MIRI filters. 
To be included in the final sample, we required sources to 1) have a SNR $>3$ in at least two MIRI bands and 2) be well-fit by a mid-IR template (see Sections \ref{sec:temp} and \ref{sec:template_fit} for a full description of the template fitting procedure). There are 911 sources with SNR $>3$ in at least one MIRI band. Of these, there are 575 (466) that have SNR $>3$ in at least two (three) bands. From the parent sample of 575 sources, 482 were well-fit by mid-IR templates and are included in the final sample. The remaining 93 sources have faint fluxes ($<1\,\mu$Jy) in the mid-IR ($\lambda>5\,\mu$m) and no unambiguous PAH features.

\subsection{Redshifts}
\label{sec:redshifts}
We cross-match our MIRI sources with the CANDELS \citep{grogin2011,koekemoer2011} EGS catalog of \citet{finkelstein2022}, which includes {\it HST}/ACS+WFC3 and {\it Spitzer}/IRAC photometry and  photometric redshifts derived with the {\tt EAZY} code \citep{brammer2008}, fitting a suite of templates to the {\it HST}+IRAC photometry. The code produces a redshift posterior probability distribution which includes the 68\% confidence intervals. The final redshift is the peak of the probability distribution. 
As discussed below, for some MIRI sources we use photometric redshifts from the CANDELS catalog of \citet{stefanon2017}, measured by fitting {\it HST} and IRAC data with various different codes. The ${z}_{\mathrm{phot}}$ is the median redshift determined through five separate codes that fit templates to the UV/optical/near-IR data.
A mere 20 sources have spectroscopic redshifts, and we use these where available.

Optical photometric redshifts can carry a degree of uncertainty if templates are not representative of the full galaxy population or if the photomery is noisy or sparse \citep{newman2022}. MIRI covers the PAH features for the majority of our galaxies, which may provide additional constraints on the redshifts due to their prominence \citep{pope2008,kirkpatrick2012}. MIRI is the first instrument with closely spaced filters allowing for the identification of PAH features with photometry alone.

To test how well fitting templates to mid-IR data can reproduce optical redshifts, we fit all galaxies with three templates described in more detail in Section \ref{sec:temp}. Two templates are a star-forming template (`MIR0.3') and an AGN template (`MIR0.7') from the \citet[hereafter, K15]{kirkpatrick2015} library. The third template is the $L_{\rm IR}=10^{10}\,L_\odot$ template from the \citet[hereafter, CE01]{chary2001} library. We include the IRAC 4.5\,$\mu$m and 5.8\,$\mu$m photometric points, along with all available MIRI data, in the fitting as these are useful in identifying the stellar bump at $\lambda\sim1.6\,\mu$m and the stellar minimum at $\lambda\sim5\,\mu$m. For all but 32 sources, the MIRI redshift agrees with the optical redshift from
\citet{finkelstein2022} within the confidence interval on the optical redshift. We therefore adopt the 
\citet{finkelstein2022} redshifts for these galaxies.
Of the 32 sources where the two redshifts do not agree, the MIRI redshifts of 24 sources agree with the optical redshifts from \citet{stefanon2017} within the confidence interval on the optical redshift. We adopt the \citet{stefanon2017} redshifts for these sources. For the remaining 8 sources, we adopt the MIRI redshifts. 

\subsection{MIPS 24\,$\mu$m Observations}
The EGS field was observed with {\it Spitzer}/MIPS as part of the {\it Spitzer}/FIDEL Legacy project\footnote{DOI: https://www.ipac.caltech.edu/doi/irsa/10.26131/IRSA410} \citep[PI M.\,Dickinson;][]{magnelli2009}. 
The typical MIPS exposure time per position within the CANDELS/CEERS areas is 14,000s, and the MIPS data analyzed here cover the CANDELS WFC3 field of $\sim$200\,arcmin$^2$, nearly 20$\times$ larger than the CEERS MIRI area. The MIPS beam size is 6$^{\prime\prime}$, causing blending of closely spaced sources within each beam, which have been carefully deblended based on {\it HST} and IRAC prior positions and fluxes. 
Photometry has been calculated with a PSF-fitting method to extract MIPS fluxes at the positions of the  {\it HST}/F160W priors from the catalog of \citet{stefanon2017}. The methodology is similar to what is described in \citet{jin2018} and \citet{liu2018}, and full details will be provided in a forthcoming paper (Le Bail et al.\,in preparation). Flux biases and uncertainties were measured using extensive simulations. 
There are $\sim$4000 MIPS sources with $S_{\rm 24}>21\mu$Jy, which is the average 3$\sigma$ flux limit in the Le Bail catalog.

Figure \ref{fig:image} illustrates the image quality of MIRI compared with IRAC and MIPS. With MIPS, the larger angular resolution leads to multiple galaxies falling within one beam. MIRI observations alleviate this challenge, enabling the detection of fainter, non-isolated sources.

\subsection{Infrared Templates}
\label{sec:temp}
To measure redshifts, estimate $L_{\rm IR}$, and search for AGN emission, we fit a suite of templates to the MIRI population (Section \ref{sec:template_fit}). These templates are the K15 MIR Library and the CE01 Library. We opt for this method as it is less computationally expensive than codes such as {\tt CIGALE}. The K15 and CE01 templates are empirically derived from infrared observations of galaxies, while {\tt CIGALE} and other spectral energy distribution (SED) fitters produce model spectra using stellar population synthesis, dust emission models, radiative transfer calculations, or energy balance requirements.

The K15 templates were empirically created from 343 24$\mu$m-selected galaxies spanning the redshift range $z\sim0.5-3.0$. Every galaxy had a {\it Spitzer}/IRS spectrum (low resolution spectroscopy) that was modeled by combining a PAH template with power-law emission attributed to a torus. This spectral modeling determined \fagn, which is the fraction of mid-IR emission attributable to the power-law component. \fagn\ spans the range 0.0-1.0 (fully star-forming to fully AGN). Additionally, all galaxies had $JHK$, {\it Spitzer}/IRAC, and {\it Herschel}/PACS and SPIRE photometry covering the full IR wavelength range.

Galaxies were sorted by \fagn\ in increments of 0.1 (that is, the pure star-forming bin would have galaxies with 0$\le$\fagn$<$0.1, while the pure AGN bin would have 0.9$\le$\fagn$\le$1.0; every other bin contains galaxies with \fagn\ indicating varying mixes of star-formation and AGN emission) to create empirical templates. From $\lambda=0.9-20\,\mu$m, the normalized photometry and spectroscopy were averaged together as a function of wavelength using a bootstrapping technique. Beyond 20\,$\mu$m, the normalized photometry was fit with a two-temperature modified blackbody.

Crucially, all galaxies in the sample used to create the K15 template library have $L_{\rm IR}>10^{11}\,L_\odot$. They are representative of the (U)LIRG population at cosmic noon.

As discussed below, lower luminosity templates are also needed to represent the MIRI population. To this end, we included the CE01 templates with $L_{\rm IR}=[2\times10^8,5\times10^8,1\times10^9,5\times10^9,1\times10^{10},5\times10^{10}]\,L_\odot$ in our fitting procedure.

CE01 used models tailored to match the ultraviolet--submillimeter SED of four galaxy prototypes: Arp 220, NGC 6090, M82, and M51. These four representative models were then divided into mid-infrared (4–20\,$\mu$m) and far-infrared (20–1000\,$\mu$m) components and fit to the ISOCAM and IRAS observations of local galaxies and (U)LIRGs in the IRAS Bright Galaxy Survey. Galaxies were sorted into luminosity bins, and the best-fitting mid-infrared and far-infrared models were averaged together to yield the final representative template SED for each luminosity bin.

\begin{figure*}[ht!]
\includegraphics[width=6.0in]{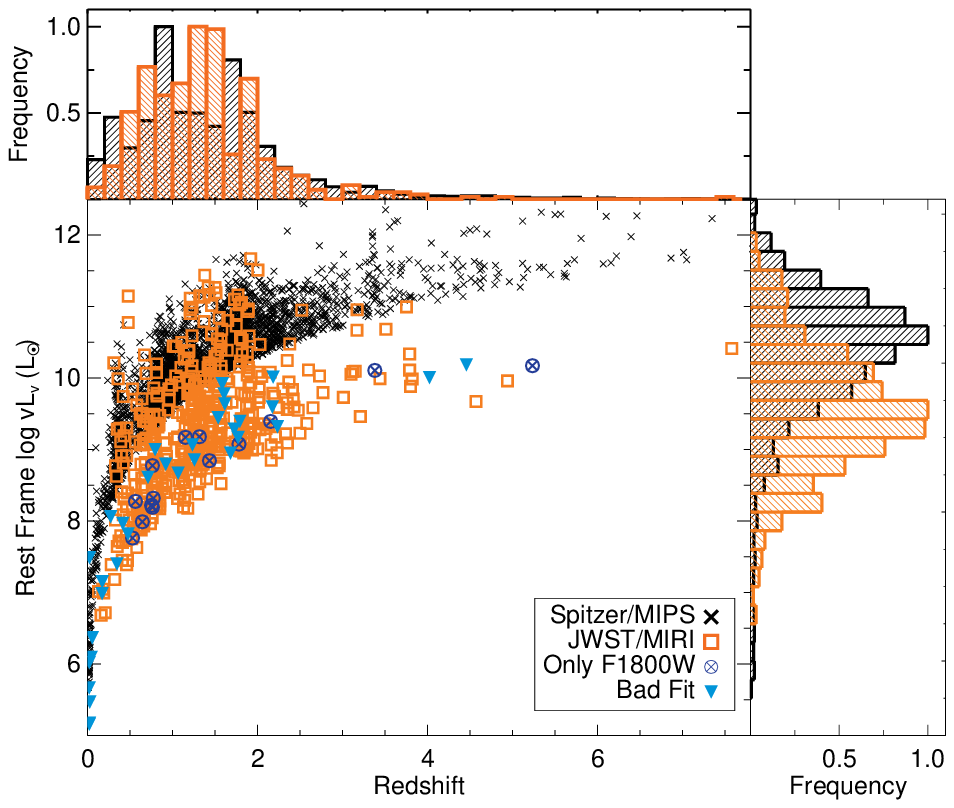}
\caption{Redshift vs.\,$\nu L_\nu$ for the full MIPS population ($\nu=24\,\mu$m; black crosses) and the MIRI popoulation ($\nu=18\,\mu$m; orange squares). The surprising lack of high-$z$ MIRI sources is attributable to the lack of bright dust features at $\lambda_{\rm rest}\leq5\mu$m and the much smaller survey area. MIRI is capable of detecting fainter sources than MIPS at all redshifts in a fraction of the time. We show the final sample (SNR$>$3 in at least 2 bands and well-fit by a mid-IR template; 385 18\,$\mu$m galaxies meet this criteria) in orange. We also show the sources which have SNR$>$3 in only the F18000W filter (14 sources; blue circles with crosses), and the sources which are not well-fit with a mid-IR template (34 sources; blue triangles). Our selection criteria is not biasing the final sample towards brighter sources. The histograms in the side panels have been arbitrarily normalized due to the stark difference in numbers of sources.
}
\label{fig:z_lir}
\end{figure*}

\section{Results}

\label{sec:results}
\subsection{Comparison with \it{Spitzer} 24\,$\mu$m}
Prior to the launch of {\it JWST}, {\it Spitzer} was the premier mid-IR observatory. 
 To better understand the new mid-IR Universe revealed through MIRI observations, we compare the properties of the CEERS {\it JWST}/MIRI galaxies (traced through the 18\,$\mu$m population) to the EGS {\it Spitzer}/MIPS population.

Figure \ref{fig:z_lir} compares the luminosity and redshift distribution of the full EGS MIPS population and the 18\,$\mu$m population. We choose to show the 18\,$\mu$m population as 18\,$\mu$m is close in wavelength to 24\,$\mu$m, F1800W was observed in all 4 pointings, and the F1800W sensitivity is better than at F2100W. There are 433 sources with SNR$\ge3$ in the F1800W filter. 

As seen in the top panel, MIPS sources have a bimodal redshift distribution with peaks at $z\sim1$ and $z\sim2$ due to PAH emission. At $z\sim1$, the 11.2\,$\mu$m and 12.7\,$\mu$m PAH complexes fall within the 24\,$\mu$m bandpass, and at $z\sim2$, the 7.7\,$\mu$m PAH feature slides in. Additionally, the 9.7\,$\mu$m silicate absorption feature falls in the bandpass at $z\sim1.5$, contributing to the dearth of detections. 
The 18\,$\mu$m distribution also has noticeable peaks where different PAH features slide into the filter. The peak around $z\sim0.5$ is attributable to the 11.2 and 12.7\,$\mu$m features. The majority of 18\,$\mu$m sources lie around $z\sim 1.3$. This is due to the 7.7\,$\mu$m complex falling in the F1800W bandpass at $z=1.3$. The 7.7\,$\mu$m complex is the broadest of the PAH features, spanning $\lambda=7.2-8.2\,\mu$m. This rest wavelength range will cause a brightening at $\lambda_{\rm obs}=18\,\mu$m from $z=1.2-1.5$. The 6.2\,$\mu$m PAH feature falls in the F1800W bandpass at $z=1.9$, also visible in the redshift distribution. 


The redshift distributions are most instructive when considered alongside 
the $\nu L_\nu$ distributions. Here, $\nu L_\nu$ refers to the restframe photometry in either the 24\,$\mu$m or F18000W filter. 
MIRI galaxies are an order of magnitude less luminous in the infrared, at all redshifts, which can be attributed to the increased sensitivity. It is interesting, and perhaps unexpected, to note the lack of MIRI detections of bright sources at $z>3$. At $z>3$, F1800W probes the faintest dust emission ($\lambda\lesssim5\mu$m), contributing to the lack of high redshift sources. In terms of numbers of high$-z$ sources, MIPS has the advantage of a much larger FOV than MIRI (12$\times$ greater). 
Uncovering a high redshift population of dusty galaxies will require a large area MIRI survey with the longest wavelength filters in order to probe the dust emission at higher redshifts.

The 18\,$\mu$m population has a very similar luminosity and redshift distribution as the 21\,$\mu$m population. There are 121 sources in pointings 1 and 2 detected with MIPS and 264 sources detected in MIRI F2100W. The 21\,$\mu$m MIRI distribution peaks at more than an order of magnitude lower flux than the MIPS distribution, attributable to the improved sensitivities of {\it JWST} compared with {\it Spitzer}, largely due to the nearly $8\times$ increase in diameter. F1800W reaches fainter flux limits than F2100W in similar exposure times, but F2100W can trace dust emission out to higher redshifts.

Figure \ref{fig:z_lir} highlights MIRI's strengths: low luminosity galaxies at cosmic noon ($z=1-2$). Here, MIRI probes a population inaccessible from {\it Spitzer}, and it does it in less than an hour. The lack of filters at $\lambda>21\,\mu$m mean that the brightest dust emission, the PAH features, are no longer detectable at $z>2$. This makes MIRI ideally suited for detecting galaxies well below the knee of the luminosity function at cosmic noon.




\begin{figure*}
    \centering
    \includegraphics[width=0.95\linewidth]{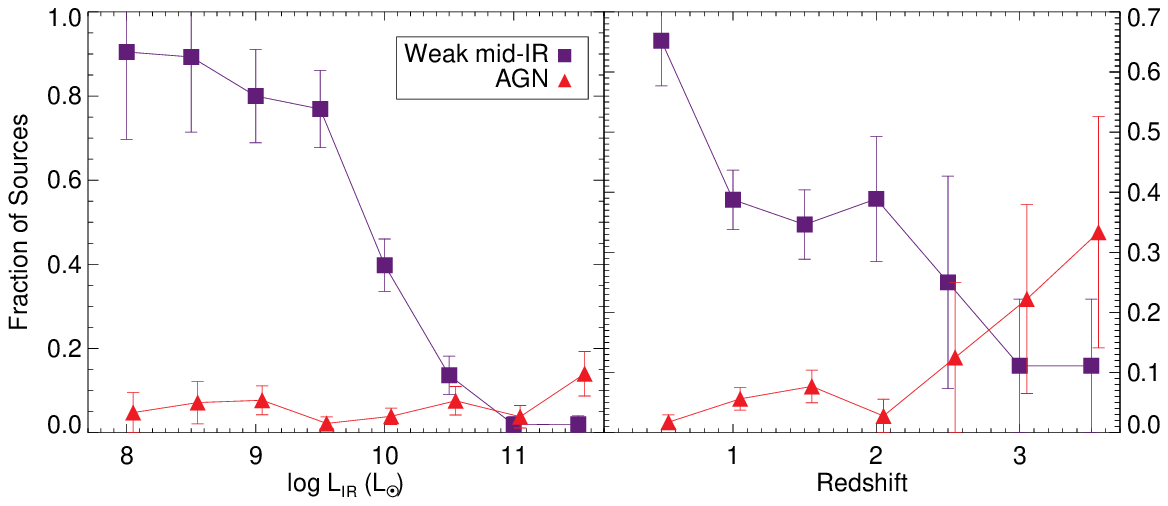}
    \caption{The fraction of galaxies that are mid-IR weak (i.e, fit with CE01 templates rather than K15; blue squares) or AGN (\fagn$>0.5$; red triangles) as a function of $L_{\rm IR}$ (Left) and redshift (Right). The small number statistics make it difficult to narrow down the true distribution of AGN with redshift, although it seems to be increasing.}
    \label{fig:agn_frac}
\end{figure*}

\subsection{Determining What Powers the Mid-IR Emission}
\label{sec:template_fit}
\citet{kirkpatrick2017} predicted that MIRI observations alone could be used to quantify AGN emission. We test that prediction by estimating AGN fraction, \fagn, by initially fitting only the template library from K15 to the MIRI data. 
Template fitting is a complementary technique to full SED decomposition. The advantages of fitting templates are that it is less computationally expensive, there are fewer parameters resulting in fewer degeneracies, and it can be performed when only a few photometric observations are available.

We fit the suite of templates to all galaxies that have two or more MIRI photometric points with SNR $>3.0$. We only include the MIRI data in this initial fitting, since this is the range that the templates are most sensitive to.  We only fit one parameter, which is the scaling of the template to the photometry. 
To assess the goodness-of-fit, we calculate a reduced $\chi^2$ parameter, and we accept all fits with reduced $\chi^2=0.5-2.0$. %
We then visually inspect the fits and overplot the {\it Spitzer}/IRAC, MIPS, and {\it HST}/WFC3 (F125W, F160W) photometry. 98\% galaxies with $L_{\rm IR} \ge 10^{11}\,L_\odot$ and 70\% of galaxies with $10^{10}\,L_\odot \le L_{\rm IR} < 10^{11}\,L_\odot$ are well-fit with the K15 templates. This is not surprising since the templates were created from IR-luminous galaxies with $L_{\rm IR}\sim10^{11}-10^{12}\,L_\odot$. Compellingly, upon visual inspection, the non-MIRI photometry follows the best-fitting template as well. 

However, the low luminosity galaxies probed by MIRI are very poorly described by the K15 template library below the mid-IR regime. Even when the reduced $\chi^2$ is acceptable, visual inspection shows that the observed near-IR emission is an order of magnitude brighter or more than the template emission.

In order to better capture the emission of lower-luminosity galaxies, we expanded the template fitting to include templates from the  CE01 Library. We fit the K15 and CE01 templates, this time including the {\it Spitzer}/IRAC 4.5\,$\mu$m data point in the fitting to help anchor the ratio of near-IR to mid-IR emission. 
We integrate each best-fit template from $8-1000\,\mu$m to estimate the galaxy's infrared luminosity. 
As stated in Section 2.1, we are able to achieve good template fits for 482 galaxies, and these galaxies comprise the final sample. Interestingly, we find that the best-fitting CE01 template $L_{\rm IR}$ does not necessarily correlate with the galaxy $L_{\rm IR}$ derived from the renormalized SED templates. In $\sim40\%$ of cases, these numbers differed by more than an order of magnitude.

In the initial fitting, where only the MIRI data points were fit with the K15 templates, a surprising 43\% of galaxies with $L_{\rm IR}<10^{11}\,L_\odot$ were classified as AGN (\fagn $>$0.5). When the 4.5\,$\mu$m data point was included, along with the CE01 templates, this number dropped to 10\%. 
Through our template fitting, we identify 28 AGN (\fagn$>$0.5) out of our final sample of 482 galaxies. For context, \citet{yang2023} fit UV-MIRI photometry with {\tt CIGALE} for our same sample, 
and they identify 31 AGN.
We compare our template fitting results with the {\tt CIGALE} decomposition 
and find good overall agreement. For 60\% of sources, the AGN fractions agree to within 10\%, and for 95\% of sources, the fractions agree to within 50\%. Broadly speaking, both manners of measuring \fagn\ will classify the same galaxies as AGN, although the exact fractions may vary. 

The K15 templates were created from dust-rich galaxies with strong PAH emission, and the ratio of mid-IR emission to near-IR emission reflects the fact that most of the galaxy's energy is being emitted in the infrared. In contrast, the lower-luminosity CE01 templates represents galaxies with less mid-IR emission compared with the strength of their near-IR emission. We label sources best fit with the CE01 templates as ``mid-IR weak" galaxies. Figure \ref{fig:agn_frac} shows how the AGN fraction and mid-IR weak fraction varies with luminosity and redshift. As expected, the mid-IR weak fraction increases sharply below $L_{\rm IR}=10^{10}\,L_\odot$. 
The fraction of mid-IR weak galaxies decreases with redshift, although this is likely driven by the flux limit of the survey.
The fraction of galaxies that are AGN increases with redshift. The increase in AGN is similar to what was found in \citet{yang2023}. 
The increase in the number of detected AGN is likely due to their enhanced emission around $5\mu$m compared with star forming galaxies, making them easier to detect in the longer wavelength MIRI filters at higher redshift.

\subsection{Average Emission of Mid-IR Selected Galaxies}

MIRI has opened a deeper window on the infrared Universe, 
particularly at cosmic noon. We explore the ratio of 
near-IR to mid-IR emission in MIRI galaxies through the 
creation of average SEDs. We sort galaxies into 
luminosity bins of $\log L_{\rm IR}=0.5\,$dex, which 
allows for at least 35 galaxies in each bin. We then 
calculate the median $L_{\rm IR}$ in each bin 
and normalize each galaxy to that value. The $L_{\rm IR}$s used to sort and normalize the galaxies are those derived by fitting the K15 and CE01 templates. We discuss $L_{\rm IR}$ further below. We remove all 
galaxies with \fagn$>0.3$. We list the properties of 
each luminosity bin in Table \ref{tbl:sed_prop}.

\begin{deluxetable*}{rcccc}
\tablecolumns{5}
\tablecaption{Properties of Luminosity Bins and SEDs}
\label{tbl:sed_prop}
\tablehead{
    \colhead{Bin} & \colhead{Sources} & \colhead{Median $\log L_{\rm IR}$\tablenotemark{a}} & \colhead{Median $z$} & \colhead{Average SED 
    $\log L_{\rm IR}$\tablenotemark{b}} \\
    \colhead{($\log L_{\rm IR}$)} & \colhead{} & \colhead{($L_\odot$)} & \colhead{} & \colhead{($L_\odot$)}}
\startdata
$<9.0$ & 50 & 8.52 [8.22,8.77] & 0.58 [0.41,0.68] & 8.44 [0.55]\\
9.0-9.5 & 54 & 9.26 [9.13,9.38] & 0.90 [0.74,1.23] & 9.02 [0.06]\\
9.5-10.0 & 79 & 9.74 [9.65,9.86] & 1.33 [0.92,1.55] & 9.53 [0.07] \\
10.0-10.5 & 77 & 10.24 [10.10,10.38] & 1.32 [0.79,1.82] & 9.88 [0.06] \\
10.5-11.0 & 43 & 10.72 [10.59,10.90] & 1.42 [1.08,1.89] & 10.29 [0.04]\\
11.0-11.5 & 39 & 11.14 [11.09,11.31] & 1.57 [1.19,2.15] & 10.66 [0.04] \\
$>11.5$ & 35 & 11.75 [11.55,11.94] & 1.53 [1.32,1.81] & 11.26 [0.04]
\enddata
\tablenotetext{a}{Median $\log L_{\rm IR}$ and redshift [with lower, upper quartiles] for sources per bin. $L_{\rm IR}$ is extrapolated for individual sources using the best-fitting K15 or CE01 template.}
\tablenotetext{b}{Average $L_{\rm IR}$ [with standard deviation] of $\log L_{\rm IR}$ values extrapolated by fitting the average SED per luminosity bin using several different template libraries.}

\end{deluxetable*}
\begin{figure*}[ht!]
    \centering
    \includegraphics[width=0.95\linewidth]{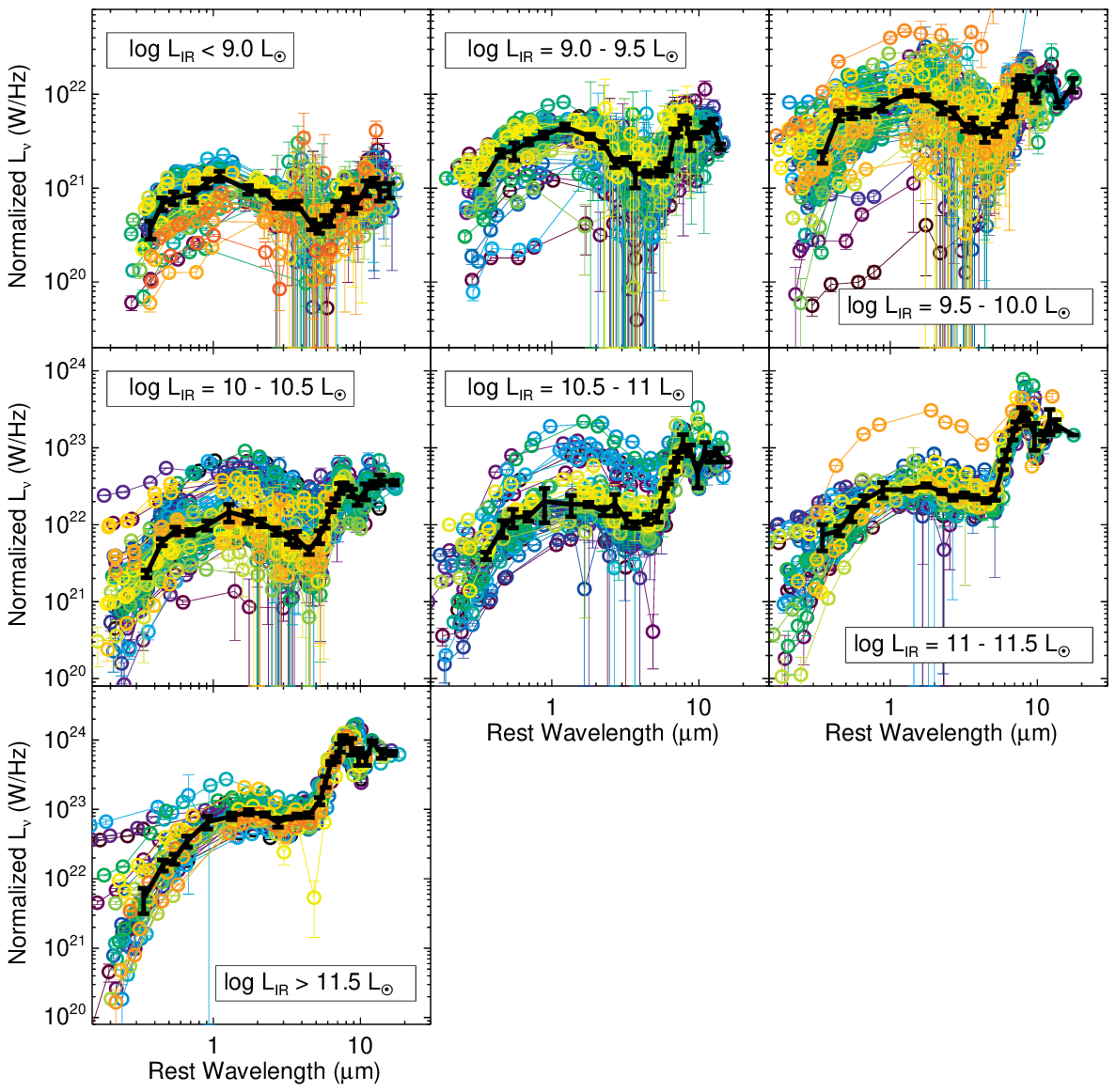}
    \caption{To create average SEDs (shown in black), we separated galaxies by luminosity and normalized to the median luminosity in each bin. Individual galaxies are plotted by color.}
    \label{fig:all_seds}
\end{figure*}

We include the {\it HST}/ACS (F606W, F814W), WFC3 (F125W, F160W), {\it Spitzer}/IRAC, MIPS 24\,$\mu$m, and MIRI photometry, redshifted to the rest-frame, in the creation of the SEDs. 
 We use a bootstrap technique where we resample the sources in each bin with replacement 1000 times. In each iteration, we calculate the median luminosity in wavelength bins of differential sizes. After 1000 iterations, we calculate the average luminosity in each bin (this is the average of the median luminosities calculated in each iteration) and the standard deviation. We list these values in Table \ref{tbl:seds}. Figure \ref{fig:all_seds} shows all of the galaxies in each luminosity bin and the resulting average SED. It is notable that the noisiest photometry in each bin occurs around $\lambda\sim5.0\,\mu$m, where the emission switches from being dominated by the older stellar population to being dominated by dust. The highest luminosity bin shows the most uniformity among individual galaxies. 

\begin{deluxetable*}{r|rr|rr|rr|rr}[ht!]
\tabletypesize{\footnotesize}
\tablecolumns{9}
\tablecaption{Average Spectral Energy Distributions}
\label{tbl:seds}
\tablehead{
    \colhead{Wavelength} & \colhead{$L_\nu$} & \colhead{Unc.} & \colhead{$L_\nu$} & \colhead{Unc.} & \colhead{$L_\nu$} & \colhead{Unc.} & \colhead{$L_\nu$} & \colhead{Unc.}\\ 
    \colhead{($\mu$m)} & \colhead{(W/Hz)} & \colhead{(W/Hz)} & \colhead{(W/Hz)} & \colhead{(W/Hz)} & \colhead{(W/Hz)} & \colhead{(W/Hz)} & \colhead{(W/Hz)} & \colhead{(W/Hz)}}
\startdata
 \multicolumn{1}{c}{} & \multicolumn{2}{|c|}{$\log L_{\rm IR}<9.0 L_\odot$} & \multicolumn{2}{c}{$\log L_{\rm IR}=9.0-9.5 L_\odot$} & \multicolumn{2}{|c}{$\log L_{\rm IR}=9.5-10.0 L_\odot$} & \multicolumn{2}{|c}{$\log L_{\rm IR}=10.0-10.5 L_\odot$}\\
 \hline
  0.335 & 3.66e20 & 7.97e19 & 1.24e21 & 1.57e20 & 2.14e21 & 2.98e20 & 2.30e21 & 2.55e20 \\
 0.454 & 6.90e20 & 1.08e20 & 2.36e21 & 2.36e20 & 5.96e21 & 7.42e20 & 6.10e21 & 6.39e20 \\
 0.544 & 7.91e20 & 1.21e20 & 2.49e21 & 5.00e20 & 6.25e21 & 8.16e20 & 8.19e21 & 9.12e20 \\
 0.663 & 8.32e20 & 1.38e20 & 2.98e21 & 3.06e20 & 6.24e21 & 5.70e20 & 8.06e21 & 7.60e20 \\
 0.913 & 1.14e21 & 1.32e20 & 3.74e21 & 2.89e20 & 7.78e21 & 1.06e21 & 9.91e21 & 1.39e21 \\
 1.330 & 1.20e21 & 1.41e20 & 4.33e21 & 3.17e20 & 1.02e22 & 7.18e20 & 1.44e22 & 3.56e21 \\
 1.740 & 9.60e20 & 9.93e19 & 3.69e21 & 2.16e20 & 9.17e21 & 7.31e20 & 1.25e22 & 2.16e21 \\
 2.243 & 8.94e20 & 4.25e19 & 2.92e21 & 2.38e20 & 7.14e21 & 9.80e20 & 1.01e22 & 1.18e21 \\
 2.750 & 6.63e20 & 5.02e19 & 1.86e21 & 1.88e20 & 6.21e21 & 7.93e20 & 7.99e21 & 9.35e20 \\
 3.220 & 6.64e20 & 5.81e19 & 1.93e21 & 1.82e20 & 4.58e21 & 4.17e20 & 6.92e21 & 1.15e21 \\
 3.722 & 6.62e20 & 6.59e19 & 1.42e21 & 4.00e20 & 4.67e21 & 8.32e20 & 6.48e21 & 1.45e21 \\
 4.462 & 4.67e20 & 5.32e19 & 1.43e21 & 9.95e19 & 3.73e21 & 6.27e20 & 4.65e21 & 5.52e20 \\
 5.284 & 3.78e20 & 4.36e19 & 1.48e21 & 1.29e20 & 4.18e21 & 6.19e20 & 7.21e21 & 1.33e21 \\
 5.797 & 4.51e20 & 6.12e19 & 1.63e21 & 2.54e20 & 5.06e21 & 5.44e20 & 1.06e22 & 1.07e21 \\
 6.310 & 4.71e20 & 5.90e19 & 2.20e21 & 2.94e20 & 6.89e21 & 1.01e21 & 1.67e22 & 1.85e21 \\
 6.784 & 6.38e20 & 7.38e19 & 3.87e21 & 4.07e20 & 7.08e21 & 1.33e21 & 2.05e22 & 2.08e21 \\
 7.269 & 6.91e20 & 7.97e19 & 4.17e21 & 3.66e20 & 1.31e22 & 1.22e21 & 3.00e22 & 4.30e21 \\
 7.635 & 8.03e20 & 1.18e20 & 5.03e21 & 4.15e20 & 1.33e22 & 2.15e21 & 3.32e22 & 1.69e21 \\
 8.259 & 8.75e20 & 1.02e20 & 4.93e21 & 4.02e20 & 1.24e22 & 1.08e21 & 2.98e22 & 1.80e21 \\
 8.683 & 7.23e20 & 1.21e20 & 3.42e21 & 6.18e20 & 1.33e22 & 1.88e21 & 2.68e22 & 2.66e21 \\
 9.731 & 7.53e20 & 1.07e20 & 3.71e21 & 2.88e20 & 8.78e21 & 6.21e20 & 2.11e22 & 2.85e21 \\
10.919 & 1.00e21 & 1.83e20 & 4.37e21 & 5.27e20 & 1.33e22 & 1.50e21 & 2.79e22 & 6.23e21 \\
12.052 & 1.12e21 & 1.35e20 & 4.75e21 & 5.47e20 & 1.39e22 & 2.67e21 & 3.22e22 & 3.94e21 \\
13.908 & 9.27e20 & 1.91e20 & 2.86e21 & 2.02e20 & 8.21e21 & 7.60e20 & 3.80e22 & 5.43e21 \\
16.462 & 9.50e20 & 1.68e20 & 2.76e21 & 1.85e20 & 1.12e22 & 1.66e21 & 3.57e22 & 2.83e21 \\
\hline
\multicolumn{1}{c}{} & \multicolumn{2}{|c}{$\log L_{\rm IR}=10.5-11.0 L_\odot$} & \multicolumn{2}{|c}{$\log L_{\rm IR}=11.0-11.5 L_\odot$} & \multicolumn{2}{|c|}{$\log L_{\rm IR} > 11.5 L_\odot$}\\
\hline
 0.335 & 3.97e21 & 4.47e20 & 7.31e21 & 2.75e21 & 5.21e21 & 2.06e21 \\
 0.454 & 8.55e21 & 2.11e21 & 8.62e21 & 1.26e21 & 1.58e22 & 2.92e21 \\
 0.544 & 1.24e22 & 3.34e21 & 1.38e22 & 3.02e21 & 1.94e22 & 3.28e21 \\
 0.663 & 1.27e22 & 1.95e21 & 2.08e22 & 3.08e21 & 3.48e22 & 6.27e21 \\
 0.913 & 2.00e22 & 9.34e21 & 2.87e22 & 6.26e21 & 6.61e22 & 1.41e22 \\
 1.330 & 1.84e22 & 5.82e21 & 2.86e22 & 1.49e21 & 7.97e22 & 8.70e21 \\
 1.740 & 1.84e22 & 1.60e21 & 3.24e22 & 1.63e21 & 9.16e22 & 8.06e21 \\
 2.243 & 1.50e22 & 1.90e21 & 2.69e22 & 1.44e21 & 8.44e22 & 4.86e21 \\
 2.750 & 1.79e22 & 5.37e21 & 2.27e22 & 1.59e21 & 6.52e22 & 9.99e21 \\
 3.220 & 1.15e22 & 2.24e21 & 2.43e22 & 1.53e21 & 7.50e22 & 2.96e21 \\
 3.722 & 1.00e22 & 1.01e21 & 2.28e22 & 1.29e21 & 8.07e22 & 5.04e21 \\
 4.462 & 1.18e22 & 1.76e21 & 2.07e22 & 1.32e21 & 8.31e22 & 6.66e21 \\
 5.284 & 1.37e22 & 2.26e21 & 3.09e22 & 4.60e21 & 1.32e23 & 1.62e22 \\
 5.797 & 2.49e22 & 4.67e21 & 5.70e22 & 7.32e21 & 2.53e23 & 4.62e22 \\
 6.310 & 4.07e22 & 3.79e21 & 1.00e23 & 1.34e22 & 4.67e23 & 2.32e22 \\
 6.784 & 5.54e22 & 7.34e21 & 1.44e23 & 2.00e22 & 5.30e23 & 4.59e22 \\
 7.269 & 8.39e22 & 9.27e21 & 2.10e23 & 2.48e22 & 7.79e23 & 1.06e23 \\
 7.635 & 1.08e23 & 2.35e22 & 2.52e23 & 4.26e22 & 1.09e24 & 7.23e22 \\
 8.259 & 9.49e22 & 4.55e21 & 2.58e23 & 3.15e22 & 9.74e23 & 7.57e22 \\
 8.683 & 9.37e22 & 1.48e22 & 2.27e23 & 4.10e22 & 8.25e23 & 2.20e23 \\
 9.731 & 5.00e22 & 1.64e22 & 1.50e23 & 5.35e22 & 5.93e23 & 1.63e23 \\
10.919 & 8.87e22 & 2.47e22 & 1.44e23 & 2.50e22 & 5.37e23 & 1.08e23 \\
12.052 & 7.64e22 & 1.17e22 & 2.25e23 & 7.95e22 & 9.58e23 & 2.87e22 \\
13.908 & 7.83e22 & 1.29e22 & 1.92e23 & 2.91e22 & 6.40e23 & 9.71e22 \\
16.462 & 6.51e22 & 0.00e00 & 1.61e23 & 9.23e21 & 6.45e23 & 4.16e22 \\
\enddata
\end{deluxetable*}

\begin{figure*}[ht!]
    \centering
    \includegraphics[width=0.95\linewidth]{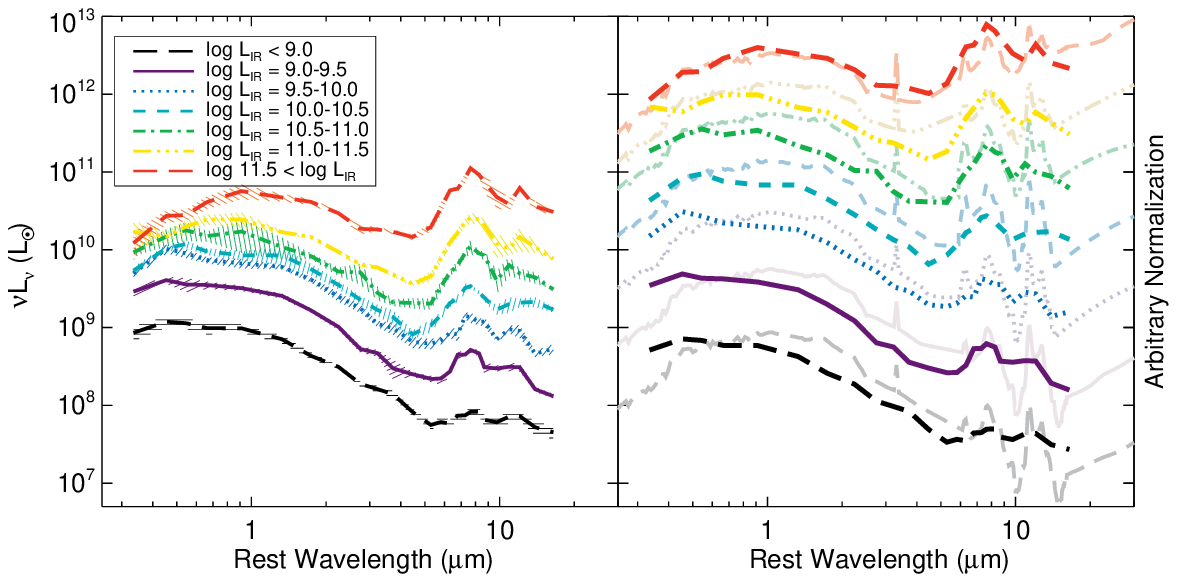}
    \caption{{\it Left panel--}Average SEDs from Figure \ref{fig:all_seds}. The PAH features clearly become stronger with increasing luminosity. At low-luminosities, the mid-IR emission is much weaker than the near-IR emission. {\it Right panel--}Average SEDs are plotted alongside the CE01 template of the same $L_{\rm IR}$. Arbitrary normalization is applied to templates of the same luminosities. The CE01 templates have more emission around $\lambda\sim1.6\,\mu$m, which may be attributable to their older stellar populations.}
    \label{fig:seds}
\end{figure*}

\vspace{-24pt}
Figure \ref{fig:seds} shows the average SEDs in the left panel. The lower luminosity SEDs clearly have more of their energy being radiated in the near-IR, which could indicate that the bulk of their star formation is unobscured. We explore this further in Section \ref{sec:discussion}. We test whether the shape of the SED changes from $z=0$ to cosmic noon by comparing with the CE01 templates in the right panel of Figure \ref{fig:seds}. We plot the CE01 template with $L_{\rm IR}$ closest to each average SED. Each CE01 template and average SED of a given luminosity have been arbitrarily normalized (the same normalization is applied to each set of a given luminosity) to allow for easier comparison. 
The average SEDs are in general remarkably similar to the CE01 SEDs, indicating that the relative amounts of near- and mid-IR emission does not evolve strongly with redshift for galaxies of the same infrared luminosity. However, it is notable that the shape of the near-IR emission does change quite strongly for most of the luminosity bins. The MIRI templates lack the strong emission seen at 1.6\,$\mu$m in the CE01 templates. This may reflect a difference in the ages of the stellar populations. The CE01 templates were created from local galaxies which have had an additional $\sim6-9$ billion years to evolve, which can make the stellar population much older and the resulting emission much stronger in the near-IR.

As a caveat--$L_{\rm IR}$ for individual sources was calculated by simply integrating the best-fit K15 or CE01 template from $8-1000\mu$m, and the $L_{\rm IR}$ for each bin listed in Table \ref{tbl:sed_prop} is the median of these values. Very few galaxies in these CEERS MIRI fields have direct far-IR detections from {\it Herschel}, so $L_{\rm IR}$ is an extrapolation from the mid-IR. If the ratio of mid-IR to far-IR emission is insensitive to redshift, then the $L_{\rm IR}$ of fainter galaxies (those missed by {\it Herschel}) can be accurately estimated from local templates.

We also calculate $L_{\rm IR}$ by fitting a suite of IR templates from CE01, \citet{dale2002}, \citet{draine2007}, and \citet{rieke2009} to each average SED (Table \ref{tbl:seds}) at $\lambda_{\rm rest}\ge 4.7\,\mu$m.
Table \ref{tbl:sed_prop} lists the average and standard deviation of the $\log(L_{\rm IR})$ values extrapolated using these four template libraries. The average $\log(L_{\rm IR})$ is about 0.5\,dex smaller than the median $\log(L_{\rm IR})$ values for individual sources per bin that were derived using the K15+CE01 templates. Except for the lowest luminosity bin, the standard deviations are small ($\leq 0.1$~dex), showing that existing SED libraries have broadly consistent mid-IR to total IR luminosity ratios, i.e., that the choice among current-generation SED templates does not have a large impact on total IR luminosities that we infer from MIRI data.  

Although the various template libraries largely agree with one another when used to extrapolate $L_{\rm IR}$ from the mid-IR emission, it is important to bear in mind that these libraries have been calibrated on local sources or infrared luminous galaxies. The ratio of mid-IR to far-IR emission can change with location on the main sequence, redshift, and metallicity \citep{elbaz2011,kirkpatrick2017b,schreiber2018}.

Direct far-IR measurements for the low-luminosity, high redshift MIRI galaxies would be necessary to verify their total luminosities.  While deep millimeter observations with ALMA, LMT/TolTEC, or other facilities can provide some constraint at wavelengths beyond the peak of bolometric emission, ultimately a new, sensitive far-IR observatory will be required to robustly measure $L_{\rm IR}$ for these galaxies.

\subsection{Color Selection}
\citet{kirkpatrick2017} used the K15 templates to predict where AGN would 
lie when combining MIRI photometric bands for color selection. 
MIRI color selection of AGN is hampered by two facts:  first, with just a few closely spaced photometric points, star-forming galaxies can mimic a power-law in the mid-IR. Therefore, MIRI color selection of AGN only has a unique solution when the redshift is known; second, if only mid-IR photometry is used, mid-IR weak galaxies can mimic the weak PAH features in AGN. We find that when plotting the real CEERS observations in the MIRI colorspaces presented in \citet{kirkpatrick2017}, the AGN regions are severely contaminated by the mid-IR weak galaxies.

\begin{figure*}
    \centering
    \includegraphics[width=0.53\linewidth]{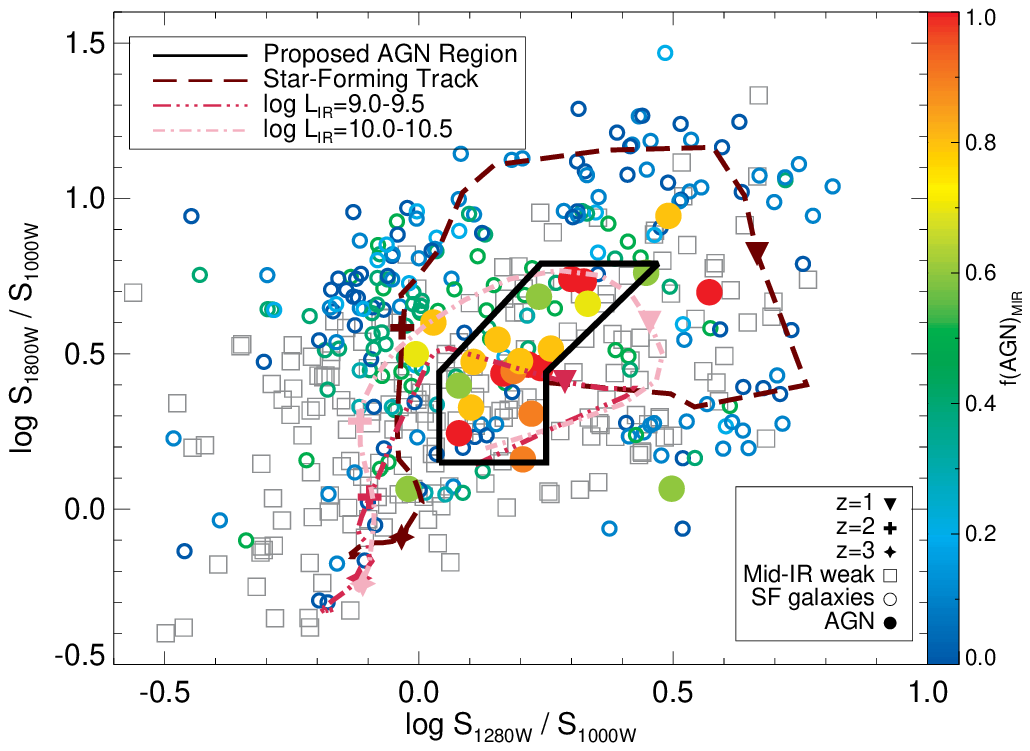}
        \includegraphics[width=0.46\linewidth]{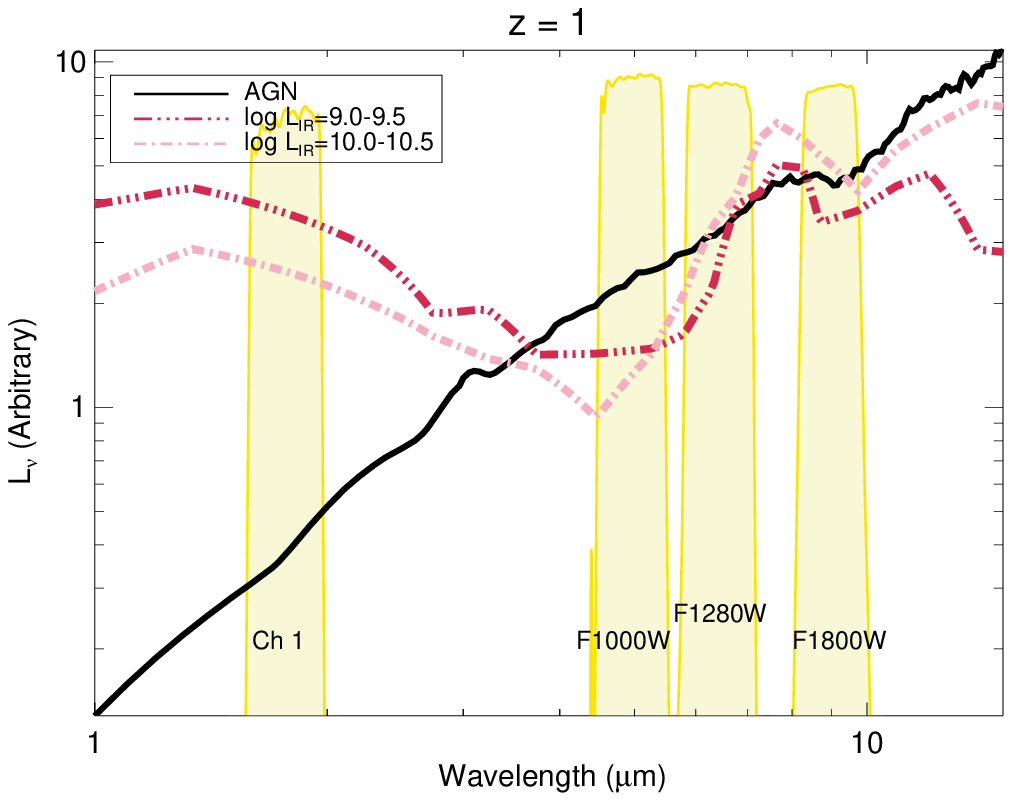}
    \caption{{\it Left panel}--MIRI color selection of AGN. Dust-rich galaxies (circles) are shaded according \fagn. Galaxies with \fagn$>0.5$ are shown as the filled circles. The mid-IR weak galaxies are the grey squares. The proposed AGN selection region (Equation \ref{eq:fubar}) is shown as the solid line. We illustrate how the dust-rich star forming galaxies (purple dashed line, K15 template) and mid-IR weak galaxies (this work, pink and light pink dot-dashed lines) move with redshift. AGN do not strongly evolve with redshift. Dust rich galaxies separate by \fagn, while the mid-IR weak galaxies contaminate the AGN regions. {\it Right panel}--mid-IR weak SEDs compared with a K15 AGN template. The IRAC Ch 1 (3.6\,$\mu$m) bandpass is shown alongside the MIRI bandbpasses. All bandpasses have been redshifted to $z=1$. At this redshift, the mid-IR weak and AGN templates have similar mid-IR colors, but the inclusion of a near-IR data point will help to separate them.}
    \label{fig:z1_col}
\end{figure*}

Instead, we present a new diagnostic that does not take into account redshift information, in order to illustrate the limited utility of using mid-IR colors alone to identify AGN. Figure \ref{fig:z1_col} combines the colors $S_{\rm F1800W}/S_{\rm F1000W}$ with $S_{\rm F1280W}/S_{\rm F1000W}$. We find that combining three filters, rather than four, does a better job at separating AGN, star-forming galaxies, and mid-IR weak galaxies.

The dust-rich galaxies (those fitted with K15 templates) are plotted as the shaded circles, while the mid-IR weak galaxies (fit with CE01 templates) are plotted as the grey squares. AGN (solid circles) lie towards the center of the colorspace, while star forming galaxies lie around the edges. 

We define an irregular hexagon to separate the AGN from the non-AGN, the vertices of which are:
\begin{align}
\label{eq:fubar}
    x=&\log(S_{\rm F1800W}/S_{\rm F1000W}) \nonumber \\
    y=&\log(S_{\rm F1280W}/S_{\rm F1000W}) \nonumber \\
    [x,y]=&[0.04,0.15], [0.04,0.44], \nonumber\\
    &[0.24,0.79], [0.47,0.79], \nonumber\\
    &[0.25,0.44], [0.25,0.15]
\end{align}
Within this region, there are 19 AGN and 48 contaminants (there are 25 AGN total in this figure). Without including redshift information, this reliability (28\%) is the best that can be achieved with MIRI colors alone.

Galaxies with PAH features move in the diagram as redshift increases. The star-forming galaxies move from the bottom of the plot at $z<0.6$ to the top right at $z\sim0.6-1.5$ (redshift track is illustrated with the purple dashed line). At greater redshifts, they move towards the middle left. 
The mid-IR weak galaxies (redshift tracks are the dot-dashed magenta and pink lines) follow the same general movement, but at lower color ratios, causing them to contaminate the region predominantly occupied by AGN. The AGN, due to their lack of PAH features, do not strongly change location due to redshift. 

As seen in the left panel of Figure \ref{fig:z1_col}, mid-IR weak galaxies contaminate the AGN region mostly at $z\sim1$. The reason for this is illustrated in the right panel. We have plotted the K15 MIR1.0 template and our $\log L_{\rm IR}=9.0-9.5$ and $\log L_{\rm IR}=10.0-10.5$ SEDs from this work. We overplot the MIRI F1000W, F1280W, and F1800W filters, shifted to the rest wavelength they would have at $z=1$. At this redshift, there simply is not a wide separation between the SEDs, so intrinsic variation between sources will cause overlap between AGN and mid-IR weak colors. It is also obvious that adding a power-law criterion (i.e. F1000W $<$ F1280W $<$ F18000W) will not help the situation as that can be true for galaxies with and without PAH features in this regime. The part of the SEDs with the largest separation is the near-IR.

\begin{figure}
    \centering
    \includegraphics[width=1.0\linewidth]{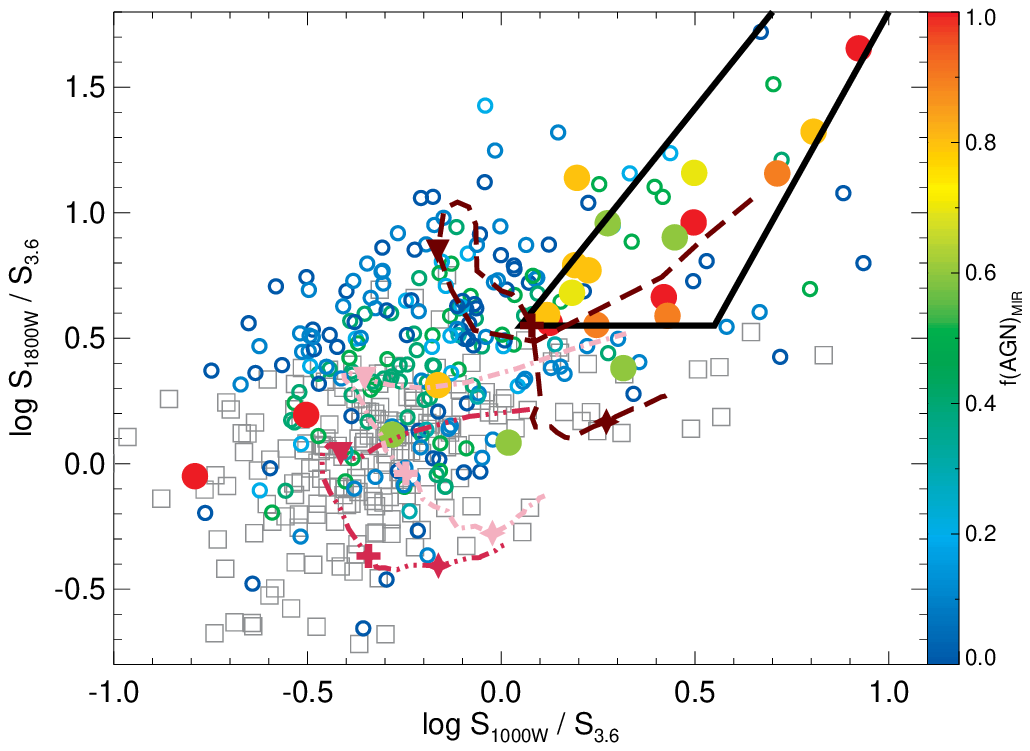}
    \caption{MIRI + near-IR color selection of AGN. Symbols and lines are the same as in Figure \ref{fig:z1_col}. 
    Dust-rich galaxies separate from mid-IR weak galaxies, in contrast with Figure \ref{fig:z1_col}. The solid lines show our proposed AGN selection region. The main contaminant in this region is dust-rich star-forming galaxies at $z<1$.}
    \label{fig:nir_col}
\end{figure}

Figure \ref{fig:nir_col} shows the filters where the three templates (plotted on the figure) have the strongest separation over the largest range of redshifts. The inclusion of a near-IR data point can help separate mid-IR weak galaxies from dust-rich galaxies. We combine IRAC and MIRI data to measure $\log S_{F1800W}/S_{3.6\,\mu{\rm m}}$ and $\log S_{F1000W}/S_{3.6}$. We use IRAC rather than NIRCam due to the uneven coverage of the CEERS observations.

The mid-IR weak galaxies (grey squares) lie in the lower left region of this colorspace, while the dust-rich galaxies lie in the upper left. The threshold $S_{\rm F1000W}/S_{3.6} \ge 0.35$ separates dust-rich star-forming galaxies from the mid-IR weak galaxies. Given the difference in luminosities between these classes of galaxies, it can also be seen as an infrared luminosity threshold. Dust-rich star-forming galaxies at $z<1$ are the main contaminant in upper right, where the AGN predominately lie. 
We recommend the following criteria for selecting AGN:
\begin{align}
x&=\log(S_{\rm F1800W}/S_{3.6}) \nonumber \\
y&=\log(S_{\rm F1000W}/S_{3.6}) \nonumber \\
y&\ge 0.55 \nonumber\\
y&\ge 2.78x-0.98 \nonumber\\
y&\le 1.92x+ 0.45
\label{eq:agn_nir}
\end{align}
Within this region, there are 15 AGN and 13 contaminants. There are 22 AGN total shown in Figure \ref{fig:nir_col}.

The MIRI-only selection (Figure \ref{fig:z1_col}) has a reliability of 28\%, while the MIR+NIR selection (Figure \ref{fig:nir_col}) has a reliability of 54\%. If we combine the mid-IR criteria (Equation \ref{eq:fubar}) with the following criteria:
\begin{align}
\label{eq:moor}
x&=\log(S_{\rm F1800W}/S_{3.6}) \nonumber \\
y&=\log(S_{\rm F1000W}/S_{3.6}) \nonumber \\
x&\ge0.55 \nonumber\\
y&\ge0.05
\end{align}
then 13 AGN are selected (out of 22) with a mere five contaminants, giving a reliability of 72\%. For AGN-candidate selection in large surveys, we recommend combining Equations \ref{eq:fubar} and \ref{eq:moor}. However, for small surveys, where computational resources are not a concern, we recommend template fitting or SED decomposition as color selection lacks either completeness or reliability.


\subsection{Number Counts}
\begin{figure}
    \centering
    \includegraphics[width=\linewidth]{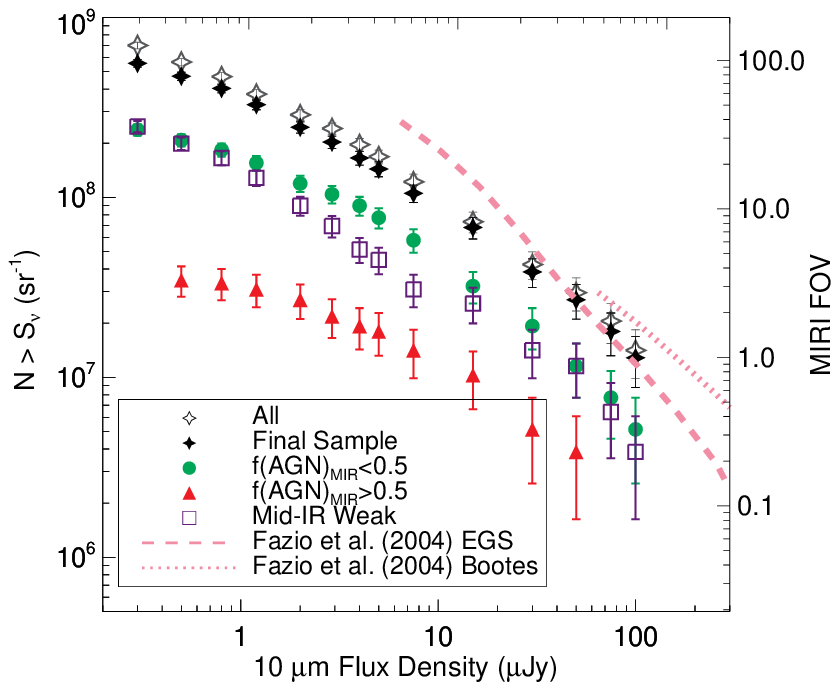}
    \caption{10\,$\mu$m cumulative number counts for all sources with SNR$>$3 in F1000W (open stars) and the final sample (filled stars). Our selection criteria mainly affected sources at the faintest F1000W fluxes. We also show the number counts for the mid-IR weak galaxies (blue squares), star-forming galaxies (\fagn$\leq 0.5$; filled green circles) and AGN (\fagn$>0.5$; red triangles). AGN predominantly have brighter F1000W fluxes. The 8\,$\mu$m number counts from \citet{fazio2004} are shown in pink. The discrepancy between the 8\,$\mu$m and 10\,$\mu$m counts may be due to the large differences in field sizes. The number of sources in one MIRI field of view is shown on the right axis.}
    \label{fig:counts}
\end{figure}

We present cumulative number counts of the 10\,$\mu$m sources in Figure \ref{fig:counts} and Table \ref{tbl:counts}. Overall, there were 573 galaxies with a $>$3$\sigma$ detection at 10\,$\mu$m (observed frame) and 469 10\,$\mu$m-detected galaxies in the final sample that have been fit with templates (these numbers differ slightly from Section 2.1 because here we are describing sources with a 10\,$\mu$m detection, rather than a detection at any wavelength). 
At the peak redshift distribution of the MIRI sources, $z=1.4$, the F1000W bandpass is probing rest-frame $4\,\mu$m emission. 

We compare the 10\,$\mu$m counts with the IRAC 8\,$\mu$m counts from \citet{fazio2004} in the EGS and Bo\"otes fields. The EGS counts match the 10$\mu$m counts at the bright end, although the Bo\"otes counts are higher. In the mid-range (10\,$\mu$Jy), the MIRI and IRAC counts also disagree. This could possibly be attributed to the much small area in the MIRI survey compared with the IRAC data. A mere 67 sources have $S_{\rm F1000W}=8.0-80.0\,\mu$Jy, which could make the underlying distribution of the parent population difficult to distinguish. A large area MIRI survey will better determine the true flux distribution of galaxies.


We also show the number counts of the mid-IR weak, star-forming (\fagn$\leq 0.5$) and AGN (\fagn$>0.5$) populations. The faint galaxy population at cosmic noon contains a large fraction of mid-IR weak galaxies (Figure \ref{fig:agn_frac}). Interestingly, both the mid-IR weak and dust-rich galaxies have approximately the same flux distribution. The mid-IR weak galaxies do not dominate the counts at faint fluxes. 
We find fewer than five AGN per MIRI pointing, illustrating that a large MIRI survey will be required to build up sufficient population statistics of infrared AGN. MIRI is capable of finding AGN in lower luminosity galaxies and galaxies at higher redshift (Figure \ref{fig:agn_frac}), but large areas are required due to the relative rarity of AGN. 

\begin{deluxetable*}{rccccc}
\tablecolumns{6}
\tablecaption{10 $\mu$m Cumulative Number Counts}
\label{tbl:counts}
\tablehead{\colhead{$S_{1000W} (\mu$Jy)} & \colhead{All Sources} & \colhead{Final Sample} & \colhead{\fagn$\leq0.5$} & \colhead{\fagn$>$0.5} & \colhead{mid-IR weak Sources}}
    \startdata
  0.3 & $6.99 \pm 0.30 \times 10^8$ & $5.58 \pm 0.27 \times 10^8$ & $2.39\pm0.18\times10^8$ & \nodata                     & $2.45 \pm 0.18 \times 10^8$ \\ 
  0.5 & $5.65 \pm 0.27 \times 10^8$ & $4.73 \pm 0.25 \times 10^8$ & $2.08\pm0.16\times10^8$ & $3.47 \pm 0.67 \times 10^7$ & $1.98 \pm 0.16 \times 10^8$ \\ 
  0.8 & $4.69 \pm 0.25 \times 10^8$ & $4.05 \pm 0.23 \times 10^8$ & $1.85\pm0.15\times10^8$ & $3.34 \pm 0.66 \times 10^7$ & $1.63 \pm 0.15 \times 10^8$ \\
  1.2 & $3.75 \pm 0.22 \times 10^8$ & $3.29 \pm 0.21 \times 10^8$ & $1.54\pm0.14\times10^8$ & $2.96 \pm 0.62 \times 10^7$ & $1.30 \pm 0.13 \times 10^8$ \\
  2.0 & $2.88 \pm 0.19 \times 10^8$ & $2.45 \pm 0.18 \times 10^8$ & $1.18\pm0.12\times10^8$ & $2.57 \pm 0.58 \times 10^7$ & $9.00 \pm 1.08 \times 10^7$ \\
  2.9 & $2.42 \pm 0.18 \times 10^8$ & $2.03 \pm 0.16 \times 10^8$ & $1.05\pm0.12\times10^8$ & $2.18 \pm 0.53 \times 10^7$ & $6.68 \pm 0.93 \times 10^7$ \\
  4.0 & $1.97 \pm 0.16 \times 10^8$ & $1.66 \pm 0.15 \times 10^8$ & $9.12\pm1.08\times10^7$ & $1.93 \pm 0.50 \times 10^7$ & $4.88 \pm 0.79 \times 10^7$ \\
  5.0 & $1.68 \pm 0.15 \times 10^8$ & $1.44 \pm 0.14 \times 10^8$ & $7.84\pm1.00\times10^7$ & $1.80 \pm 0.48 \times 10^7$ & $4.37 \pm 0.75 \times 10^7$ \\
  7.5 & $1.22 \pm 0.13 \times 10^8$ & $1.05 \pm 0.11 \times 10^8$ & $5.78\pm0.86\times10^7$ & $1.41 \pm 0.43 \times 10^7$ & $3.08 \pm 0.63 \times 10^7$ \\
 15.0 & $7.33 \pm 0.97 \times 10^7$ & $6.81 \pm 0.94 \times 10^7$ & $3.21\pm0.64\times10^7$ & $1.03 \pm 0.36 \times 10^7$ & $2.57 \pm 0.58 \times 10^7$ \\
 30.0 & $4.24 \pm 0.74 \times 10^7$ & $3.86 \pm 0.70 \times 10^7$ & $1.93\pm0.50\times10^7$ & $5.14 \pm 2.57 \times 10^6$ & $1.41 \pm 0.43 \times 10^7$ \\
 50.0 & $2.96 \pm 0.62 \times 10^7$ & $2.70 \pm 0.59 \times 10^7$ & $1.16\pm0.39\times10^7$ & $3.86 \pm 2.23 \times 10^6$ & $1.16 \pm 0.39 \times 10^7$ \\ 
 75.0 & $2.06 \pm 0.51 \times 10^7$ & $1.80 \pm 0.48 \times 10^7$ & $7.71\pm3.15\times10^6$ & \nodata                     & $6.43 \pm 2.87 \times 10^6$ \\
100.0 & $1.41 \pm 0.43 \times 10^7$ & $1.29 \pm 0.41 \times 10^7$ & $5.14\pm2.57\times10^6$ & \nodata                     & $3.86 \pm 2.23 \times 10^6$ \\
\enddata
\end{deluxetable*}

\vspace{-24pt}
\section{Discussion}
\label{sec:discussion}

\subsection{Where are All the AGN?}
\citet{kirkpatrick2017} predicted finding a significant population of faint AGN at cosmic noon by extrapolating down the luminosity function based on the number of AGN observed in IR luminous galaxies.
The MIRI observations have revealed significantly fewer 
AGN at cosmic noon and high redshift than predicted in K17. This  
can be attributed to two reasons: AGN may be harder to identify than expected or AGN may by intrinsically low luminosity in lower mass galaxies. Distinguishing between the scenarios will require larger MIRI surveys in conjunction with spectroscopic or X-ray surveys.

The most luminous, unobscured AGN exhibit power-law 
emission in the near-IR as well as the mid-IR. Heavily 
obscured AGN, however, may not be visible in the near-IR \citep[and references therein]{hickox2018}. The emission from the hottest regions of the torus will be reabsorbed by subsequent dust layers and re-emitted at longer wavelengths. This reprocessing weakens the emission at shorter infrared wavelengths. The obscured fraction of AGN rises with increasing redshift \citep{peca2023,yang2023}, which means that relying on rest-frame $\lambda < 5\,\mu$m photometric emission will likely miss a significant fraction of AGN beyond $z>3$, when the MIRI filters no longer cover the mid-IR.
 Heavily obscured AGN may have a visible stellar bump, arising from the host galaxy, although power-law emission should still be visible in the mid-IR \citep{lyu2022}. In Figure \ref{fig:nir_col}, there are  AGN that lie in the mid-IR weak region, outside of the proposed AGN selection area, indicating the diversity of near-IR emission exhibited by AGN. AGN with $\log (S_{1800W}/S_{3.6}) < 0.5$ may be more obscured. Alternately, these could belong to a class of ``hot-dust poor'' Type 1 AGN, which exhibit relatitively weak near-IR emission \citep{hao2010,lyu2017b}. The cause of the weak torus in these sources is unknown, but they have been observed out to $z=6$.

Many galaxies show a mix of power-law AGN emission and PAH emission in the mid-IR. In practice, this manifests as weaker PAH features \citep{pope2008,sajina2012,kirkpatrick2012}.
 Low luminosity AGN that are accreting at high Eddington ratios will have power-law emission from a torus \citep{mason2012}, but this emission may be outshone by the galaxy's star forming component. In low luminosity galaxies, weak PAH features may be due to swamping by an AGN, or they may be intrinsically weak due to the galaxy's dust content, as seen in Figure \ref{fig:seds}. Therefore, at low luminosity, weak PAH features are not a reliable means of identifying AGN. Additionally, low luminosity AGN that accrete at low Eddington ratios may lack a torus altogether \citep{yuan2014}. Intrinsically low luminosity AGN ($L_X<10^{42}\,$erg/s) may be difficult or even impossible to identify with photometric information alone.

In both cases (high obscuration or low luminosity), there are promising ways forward. \citet{hatcher2021} used mid-IR colors to identify AGN candidates that were not X-ray detected in the COSMOS field. The authors performed X-ray stacking to obtain the average emission of the AGN candidates and found that, indeed, as a population, the AGN candidates had X-ray signatures indicative of hosting a low luminosity AGN. X-ray stacking remains a promising way to measure the average AGN emission of a population. Specific to {\it JWST}, nebular line diagnostics will also be useful \citep{backhaus2022}, potentially moreso than looking for broad line emission, as low luminosity AGN may lack a broad line region \citep{elitzur2014}. AGN emit multiple high ionization nebular lines, including [Ne{\sc v}], [Ne{\sc vii}], [O{\sc iv}] \citep{cleri2023a,cleri2023b,negus2023}. Mid-IR spectroscopy may be the most reliable method for identifying obscured or low luminosity AGN \citep{petric2011,bonato2017,stone2022}. 

With photometry alone, headway could possibly be made by spatially decomposing the galaxies into an inner and outer region. For the most massive galaxies, this may be possible, but as is clear from Figure \ref{fig:image}, most MIRI galaxies are small in size. This is likely due to the fact that MIRI is probing lower mass galaxies. Finally, photometric or spectroscopic identification of AGN may be made more reliable through machine learning algorithms \citep{holwerda2021,poleo2023}. Machine learning is capable of combining more photometric and color information than is possible with simple color selection. Machine learning can also take into account redshift information, which is useful in identifying the PAH features. 

While identification challenges are a likely culprit for not finding more AGN at cosmic noon in our MIRI sample, there is also the possibility that bolometrically luminous AGN do not exist in large quantities in lower luminosity ($L_{\rm IR}<10^{11}\,L_\odot$) galaxies. In isolated, low-luminosity galaxies, there may not be a large enough gas supply being funnelled to the center of the galaxy to fuel black hole growth at high enough Eddington rates to be energetically dominant over the star formation and therefore detectable. Alternately, but similarly problematic, is the role that AGN duty cycle may play. AGN in low luminosity galaxies may accrete at large enough Eddington ratios to be detectable, but they may only do so for a short amount of time. If we make the simplifying assumption that all galaxies host detectable AGN at some point in their life, then the left panel of Figure \ref{fig:agn_frac} indicates that the duty cycle of these AGN is $<10\%$ of the star-forming time. In this scenario, we will never observe large amounts of AGN in low luminosity galaxies because they are in their ``dormant'' phase.

\subsection{PAH Emission}
PAH emission arises at the edges of star-forming regions, and globally, the integrated PAH emission of a galaxy has been shown to correlate with its SFR. The link between PAH emission and star formation naturally leads to a correlation with $L_{\rm IR}$, as $L_{\rm IR}$ is largely attributed to dust heating by stars formed within the past 100\,Myr. PAH emission is frequently parameterized by the luminosity of the 6.2\,$\mu$m feature, $L_{6.2}$. The ratio $L_{6.2}/L_{\rm IR}$ is observed to decrease with increasing $L_{\rm IR}$. Locally, this decrease is most evident in ULIRGs ($L_{\rm IR}\geq10^{12}\,L_\odot$). The origin of decreasing PAH emission is debated, but one possibility is that the geometry of star-forming regions is changing. In ULIRGs, star formation predominantly occurs in a large, obscured nuclear starburst. This increases the volume of dust (increasing $L_{\rm IR}$) while decreasing the surface area (decreasing $L_{6.2}$). In contrast, local LIRGs ($L_{\rm IR}=10^{11}-10^{12}\,L_\odot$) are more similar to normal star-forming galaxies in that they have many star-forming regions spread throughout the galaxy.

Intriguingly, the ratio of $L_{6.2}/L_{\rm IR}$ is higher in ULIRGs at cosmic noon compared with local ULIRGs, indicating changing conditions of star formation \citep{pope2013,kirkpatrick2014,mckinney2020}. Studies of $L_{6.2}/L_{\rm IR}$ in galaxies beyond the local universe are rare, as previously {\it Spitzer}/IRS was the only instrument capable of measuring $L_{6.2}$, and only in the brightest galaxies.

We take a first look at estimating $L_{6.2}/L_{\rm IR}$ in galaxies at $z>0$ using {\it JWST} down to $L_{\rm IR}$ an order of magnitude lower than what was possible with {\it Spitzer}. The MIRI filters cover the 6.2\,$\mu$m feature in different redshift ranges. We identify the redshift ranges for each filter where the 6.2\,$\mu$m feature is isolated, and the rising continuum beyond 6.3\,$\mu$m does not dominate the emission. We find the following redshift ranges suitable:
\begin{align}
\mbox{F770W: }  & 0.20<z<0.28\\
\mbox{F1000W: } & 0.58<z<0.67\\
\mbox{F1280W: } & 1.04<z<1.11\\
\mbox{F1500W: } & 1.39<z<1.47\\
\mbox{F1800W: } & 1.87<z<1.95
\label{eq:z}
\end{align}
We do not include F2100W as we do not have any sources in a suitable redshift range.

For each photometric point, we first estimate how much of the broadband photometry is potentially due to the 6.2\,$\mu$m emission feature using a local sample with mid-IR spectroscopy. The 5MUSES sample was observed with {\it Spitzer}/IRS and have published $L_{6.2}$ and $L_{\rm IR}$ measurements in \citet{kirkpatrick2014}. That work contains 11 star-forming galaxies spanning $z=0.06-0.24$, $\log L_{\rm IR}=10.79-11.63\,L_\odot$, and a range of $L_{6.2}/L_{\rm IR}$. The average (standard deviation) of $L_{6.2}/L_{\rm IR}$ is 0.00456 (0.00181) for the 11 5MUSES galaxies.

\begin{figure}
\includegraphics[width=\linewidth]{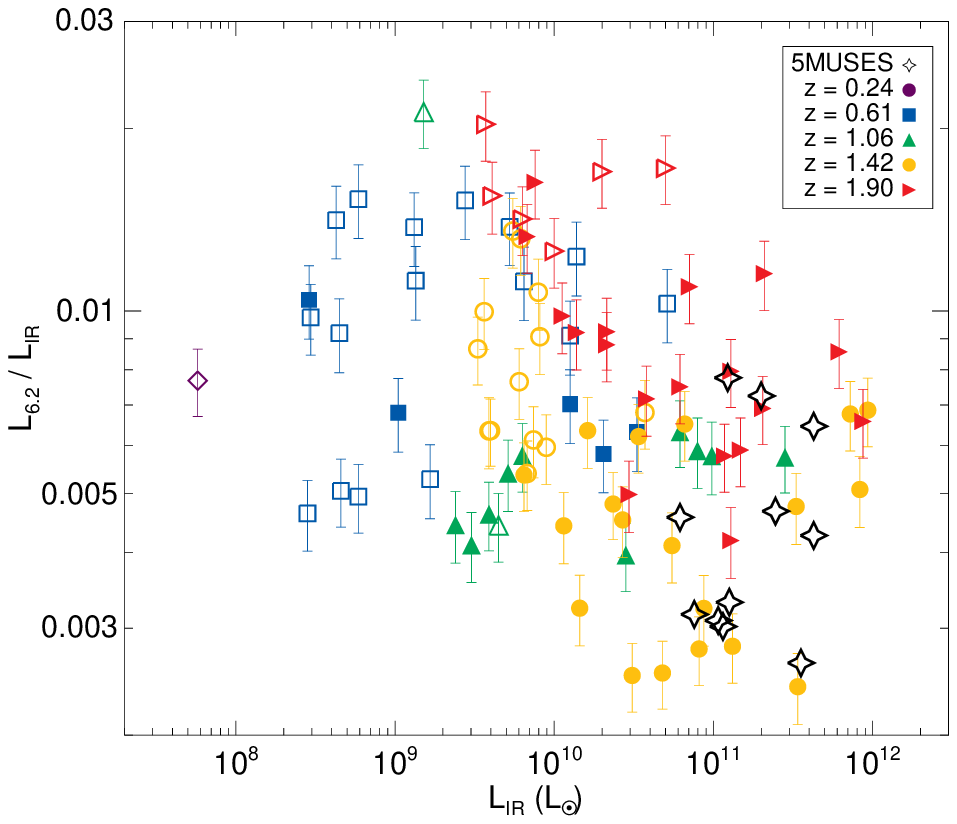}
\caption{$L_{6.2}/L_{\rm IR}$ vs. $L_{\rm IR}$ for the MIRI galaxies falling in the redshift ranges in Equation \ref{eq:z}. $L_{6.2}$ has been estimated by calculating what percentage of the photometric point could be due to the 6.2\,$\mu$m feature using the 5MUSES sample 
of star-forming galaxies from \citet[black stars]{kirkpatrick2014}. The mid-IR weak galaxies are shown as the open symbols, while the filled symbols correspond to the dust rich galaxies. Lower luminosity galaxies have higher ratios of $L_{6.2}/L_{\rm IR}$, in line with local trends. Interestingly, the higher redshift sources seems to have higher $L_{6.2}/L_{\rm IR}$, which could indicate a shift in the star-forming conditions at higher redshift. Spectroscopy is required to accurately measure $L_{6.2}$ and confirm this trend.}
\label{fig:pah}
\end{figure}

We shift the 11 5MUSES {\it Spitzer}/IRS spectra to the redshift of each MIRI galaxy and convolve with the appropriate MIRI transmission filter to calculate $\nu L_\nu$, or the photometric luminosity in a given MIRI bandpass. We use the published $L_{6.2}$ value and calculate $L_{6.2}/\nu L_\nu$ for all 11 5MUSES galaxies. We take the mean and standard deviation of $L_{6.2}/\nu L_\nu$ for the 11 galaxies as a scaling factor and associated uncertainty. We then multiply the MIRI photometry of each galaxy in our sample by this scaling factor to estimate $L_{6.2}\,[L_\odot]$ and an uncertainty. Figure \ref{fig:pah} shows $L_{6.2}/L_{\rm IR}$ as a function of $L_{\rm IR}$ for the MIRI sources. This is a conservative approach that marginalizes over uncertainty in the underlying SED based on what we expect from $z<0.2$ galaxies.  A caveat to this approach is that $L_{\rm IR}$ has been determined by fitting templates to all the MIRI and IRAC photometry, which will include the data point used to calculate $L_{6.2}$. While these calculations are somewhat independent, spectroscopy is needed to fully uncouple them.

Figure \ref{fig:pah} shows the same trend that is observed locally--$L_{6.2}/L_{\rm IR}$ decreases with increasing $L_{\rm IR}$ for a fixed redshift bin. This decrease starts to happen noticeably around $L_{\rm IR}=10^{10}\,L_\odot$. This is the same luminosity where the near-IR and mid-IR photometry begins to be best fit by dusty star-forming templates from K15, rather than mid-IR weak templates from CE01. The conditions of star formation may be significantly different in these infrared luminous sources. We show the mid-IR weak sources as the open symbols, and they have higher $L_{\rm 6.2}/L_{\rm IR}$ ratios. The decrease in $L_{\rm 6.2}/L_{\rm IR}$ with increasing $L_{\rm IR}$ is then driven by a shift from predominantly mid-IR weak sources to dust-rich sources.

Figure \ref{fig:pah} also hints at an increase in $L_{\rm 6.2}/L_{\rm IR}$ at higher redshifts (red triangles compared with the yellow circles), similar to what has been measured in sources with {\it Spitzer}/IRS spectroscopy \citep{pope2013}. This shift could indicate that the geometry and physics of star formation is changing with increasing lookback time. For example, higher $L_{6.2}/L_{\rm IR}$ ratios at earlier times could be a byproduct of the higher gas fractions in high redshift galaxies. Indeed, \citet{cortzen2019} demonstrate that PAHs are good tracers of the cold molecular gas from $z\sim0-2$, and galaxies at high redshift are increasingly gas-dominated which would increase $L_{6.2}$ relative to the dust-obscured SFR measured by $L_{\rm IR}$. Alternatively, \citet{mckinney2020} show that the offset in 
$L_{6.2}/L_{\rm IR}$ for fixed $L_{\rm IR}$ between $z\sim0$ and $z\sim2$ galaxies disappears when normalizing $L_{\rm IR}$ by the IR size. This indicates that the geometry of distributed star formation across these galaxies also plays an important role in setting the $L_{6.2}/L_{\rm IR}$ ratio. Of course, the reader should bear in mind two caveats--$L_{6.2}$ has been estimated from photometry and $L_{\rm IR}$ has been extrapolated from near- and mid-IR photometry. Robustly testing how the PAH fraction changes with luminosity and redshift will require {\it JWST} spectroscopic observations and a new far-IR telescope.

\subsection{The Nature of Mid-IR Weak Sources}
We designated sources as mid-IR weak or dust-rich based on which set of templates fit them best. Mid-IR weak sources have a much larger near-IR/mid-IR emission ratio.
The mid-IR weak status has a few potential explanations. First, these galaxies may be dust-poor, so the mid-IR emission overall is weaker than the near-IR emission, but the dust grain size distribution (i.e.\,PAH fraction) is similar to the dust-rich galaxies. On the other hand, these galaxies may be metal-poor, so that the PAH features are weaker. 
Finally, weaker mid-IR emission may be caused by a change in dust heating mechanisms. In the local Universe, spatially resolved modeling of the mid- and far-IR emission show that the temperature of dust in lower mass galaxies is lower due to the dusty disk being more extended than the stellar disk \citep{trewhella2000,dalcanton2004,xilouris2004, holwerda2009,holwerda2012}. The weak mid-IR emission is then attributable to most of the dust in low mass galaxies being colder. Morphological studies with MIRI and NIRcam may be able probe the extent of the dust and stars in $z\sim1-2$ galaxies \citep{magnelli2023,shen2023}. 

The high $L_{6.2}/L_{\rm IR}$ ratios in the mid-IR weak galaxies (Figure \ref{fig:pah}) hint that decreasing PAH emission (and hence low metallicity) is not to blame for the changing near-IR/mid-IR ratios. We also test this by fitting the average SEDs (Table \ref{tbl:seds}) with the \citep{draine2007} dust models, in which the spectra are fit with models heated by radiation fields of different strengths, for mixtures of amorphous silicate and graphite grains,
including varying amounts of PAH particles. We find that all but two of the SEDs have the exact same $q_{\rm PAH}=4.58$ as measured by these models. The $\log L_{\rm IR}<9.0$ and $\log L_{\rm IR}=10.0-10.5$ SEDs have $q_{\rm PAH}=2.50$. These results imply that the PAH spectrum is quite homogeneous for all of our galaxies, although the PAH emission only rises above the 1.6\,$\mu$m stellar bump for galaxies at higher $L_{\rm IR}$. This is in fact different from how models such as CE01 or \citet{rieke2009} behave, as the PAH features in those models become more obscured by a dust continuum at higher $L_{\rm IR}$ values. In contrast, the K15 templates have much stronger PAH features than local templates of the same luminosities.

If we take Figure \ref{fig:pah} at face value, the interpretation of the mid-IR weak sources becomes that their overall dust emission is weak, but they are not metal poor. The similar $q_{\rm PAH}$ values point to similar amounts of very small grains regardless of luminosity. The higher $L_{\rm 6.2}/L_{\rm IR}$ in mid-IR weak galaxies points to a lack of far-IR emission, which would rule out the interpretation that their dust is colder. Future far-IR observations would provide interesting insights into the ISMs of this relatively unexplored population.

\subsection{How do MIRI SFRs correlate with Optical SFRs?}
In dust rich galaxies, infrared star formation rates (SFRs) are more reliable than optical SFRs since much of the optical emission is attenuated \citep{kennicutt1998,kennicutt2012,calzetti2007,boquien2016}. It is therefore tempting to use IR observations to calculate SFRs for all infrared-detected galaxies. However, MIRI observations probe further down the luminosity function at cosmic noon than previous IR telescopes, and most CEERS MIRI sources are undetected by the available {\it Herschel} observations. This can introduce significant uncertainty in the calculation of $L_{\rm IR}$ as the ratio of far-IR to mid-IR emission has intrinsic scatter and can also depend on factors such as metallicity and whether galaxies are on the main sequence \citep{elbaz2011}. In this subsection, we compare previously calculated optical SFRs with SFRs estimated by our simplistic template fitting. 
The purpose of this section is to provide a first look at how well these SFRs compare, and whether different ways of estimating SFRs give similar answers. 
An in-depth analysis of MIRI-based SFRs, including new SFR calibrations, will be discussed in Roynane et al.\,(2023, in prep). 

\begin{figure}
    \centering
    \includegraphics[width=\linewidth]{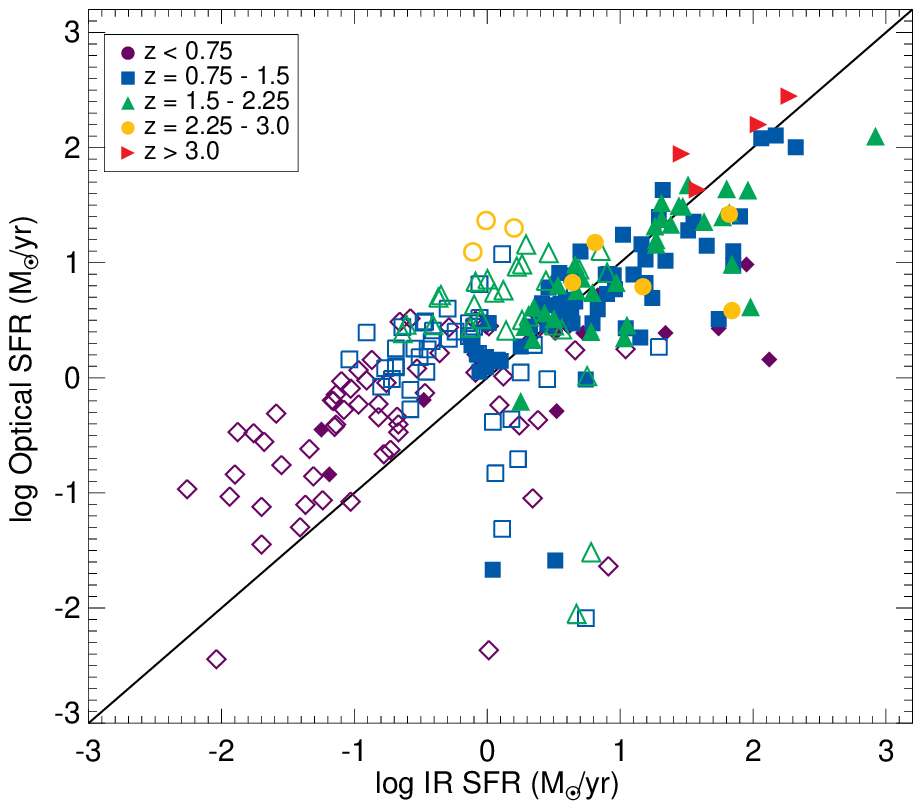}
    \caption{Optical vs.\,IR star formation rates, where symbol color and shape corresponds to redshift. Mid-IR weak galaxies are shown as the open symbols. At lower luminosities, the bulk of star formation is unobscured, which is reflected by the optical SFRs being systematically higher than the IR SFRs. This is mainly an issue for the mid-IR weak galaxies.}
    \label{fig:SFR}
\end{figure}

Figure \ref{fig:seds} shows how the near-IR and optical emission increases at lower infrared luminosities, indicating that more of the star formation is unobscured. In the bottom panel of Figure \ref{fig:SFR}, we compare IR and optical SFRs for our MIRI sample, to examine when the two indicators are in disagreement.
We calculate the infrared SFRs using the equation:
\begin{equation}
    {\rm SFR}\,[M_\odot/{\rm yr}] = 1.59\times 10^{-10} * L_{\rm IR}\,[L_\odot]
\end{equation}
which assumes a Kroupa IMF \citep{murphy2011,kroupa2001}. 

The optical star formation rates come from the CANDELS catalogs of \citet{stefanon2017}, following methods described there (see also \citet{dahlen2013} and \citet{santini2015}). 
Briefly, the star formation rates are derived by fitting the CANDELS UV/Optical
photometry in 10 different ways, each fit using a different code,
priors, grid sampling, and star formation histories. The
final value is the median from the different fits. Optical SFRs are sensitive to the photometric redshift of the galaxy. 
The optical SFRs from \citet{stefanon2017} were computed using photometric redshifts from that catalog, whereas for CEERS MIRI galaxies we also use photometric redshifts from \citet{finkelstein2022} (see Section~\ref{sec:redshifts}). For this reason, we keep only those sources where the photo-z's from \citet{stefanon2017} and \citet{finkelstein2022} agree within $\Delta z < 0.15$.
Similarly, we also remove any sources where the MIRI redshift and the optical redshift differ by more than 0.15. The remainder of the sample is shown in Figure \ref{fig:SFR}.

There is in general a one-to-one correlation between the optical and IR SFR, and there is no trend away from the one-to-one line with increasing \fagn. 
The disagreement is strongest for a handful of mid-IR weak galaxies, where the optical SFRs greatly underpredict the IR SFRs. At $L_{\rm IR} < 10^{10}\,L_{\rm \odot}$, the optical SFRs are nearly universally higher than IR SFRs. IR SFRs may underpredict the true SFR at low luminosities, when the bulk of star formation is unobscured. Interestingly, Figure \ref{fig:SFR} shows that the disagreement between optical and IR SFRs depends on whether a galaxy is classified as mid-IR weak or dust-rich, rather than on luminosity or redshift. For the mid-IR weak galaxies, the optical SFRs are higher, while for the dust-rich galaxies, the IR SFRs are higher. Dust-rich galaxies at the same redshifts or luminosities as mid-IR weak galaxies lie closer to the one-to-one line.

Optical SFRs themselves carry a high degree of uncertainty as there are many degeneracies that plague this portion of a galaxy's SED, including stellar ages and dust attenuation \citep{leja2019,bell2003}. This comparison between SFR indicators is for optical- and IR-based SFRs calculated with template-fitting to a few photometric data points. Roynane et al.\,2023 (in prep.) compares mid-IR-based SFRs with UV-based SFRs and finds a much stronger agreement. The most accurate estimations of SFRs will likely come from combining either UV or spectroscopic indicators, such as H$\alpha$, with mid-IR estimators \citep{murphy2011,kennicutt2012}.

\section{Conclusions}
We have examined the demographics of the CEERS MIRI survey, which comprises four pointings in the EGS field covering the F770W-F2100W filters. 
 We summarize our findings below.
\begin{itemize}
    \item {\bf Comparison with the {\it Spitzer} 24\,$\mu$m population:} MIRI sources peak at $z\sim1.3$ and lack many high-redshift detections seen in the MIPS 24\,$\mu$m population. This is due both to MIRI's smaller field of view and to MIRI's lower wavelength filters (which are more sensitive), no longer tracing dust beyond $z>3$. 
    A large area MIRI survey will be required to build up a large high-redshift population. However, in much shorter integration times, MIRI probes an order of magnitude fainter than MIPS due to the increased sensitivity of {\it JWST}. MIRI is best at identifying galaxies with $L_{\rm IR}<10^{10}\,L_\odot$ at cosmic noon ($z=1-2$).

    \item {\bf Mid-IR weak galaxies:} Until now, with the improved mid-IR coverage and sensitivity of {\it JWST}, it was impossible to know exactly what the mid-IR emission of $L_{\rm IR}<10^{10}\,L_\odot$ galaxies at cosmic noon looked like. We fit the MIRI and IRAC 4.5\,$\mu$m photometry with a suite of dusty star forming templates, AGN templates, and templates created from local low-luminosity galaxies. We find that inclusion of the 4.5\,$\mu$m data point is essential to discriminate between AGN and ``mid-IR weak'' galaxies, as their mid-IR emission can look very similar. It is only by comparing the mid-IR emission with rest-frame near-IR emission that one can distinguish between a mid-IR power-law due to an AGN torus and weak mid-IR emission due to an intrinsically low-luminosity galaxy. We find that mid-IR weak galaxies dominate the MIRI population at $L_{\rm IR}<10^{10}\,L_\odot$. In these galaxies, the SED is dominated by near-IR emission, indicating that the star-formation is predominantly unobscured. For that reason, we recommend that infrared SFRs should be used with extreme caution.

    \item {\bf AGN identification with colors:} The AGN fraction increases with $L_{\rm IR}$, which may reflect an increasing accretion efficiency, or it may reflect the challenges in separating AGN from mid-IR weak galaxies at lower luminosities. Spectroscopic surveys are necessary to distinguish between the two scenarios. We find that AGN can be reliably identified by combining two color selections: $S_{1800W}/S_{1000W}$ v.\ $S_{1280W}/S_{1000W}$ and $S_{1800W}/S_{3.6}$ v.
    $S_{1000W}/S_{3.6}$. 
    The 10\,$\mu$m number counts reveal that AGN are extremely rare (a handful appear in each MIRI pointing), so large area MIRI surveys are required to build up sufficient samples of mid-IR AGN to study supermassive black hole growth at cosmic noon.
    
\end{itemize}
\vspace{-4pt}
MIRI is revealing many interesting properties of galaxies at cosmic noon, but {\it JWST} results are hampered by it's small field of view and it's lack of longer wavelength filters. All the results in this paper will be strengthened with the upcoming MEGA survey (PI Kirkpatrick, Cycle 2 GO-3794) which will observe the EGS field with an additional 26 MIRI pointings. A new far-IR telescope that matches {\it JWST}'s sensitivity would also allow for the unambiguous measurement of $L_{\rm IR}$.

\vspace{12pt}

We thank the anonymous referee for the time and energy they spent improving the quality of this paper.
We gratefully acknowledge support from NASA grants
JWST-ERS-01345 and JWST-AR-02446. This work was based on observations made with the NASA/ESA/CSA James Webb Space
Telescope.
\label{sec:conclusions}

\bibliographystyle{aasjournal}
\bibliography{references}

\begin{thebibliography}{}
\expandafter\ifx\csname natexlab\endcsname\relax\def\natexlab#1{#1}\fi
\providecommand{\url}[1]{\href{#1}{#1}}
\providecommand{\dodoi}[1]{doi:~\href{http://doi.org/#1}{\nolinkurl{#1}}}
\providecommand{\doeprint}[1]{\href{http://ascl.net/#1}{\nolinkurl{http://ascl.net/#1}}}
\providecommand{\doarXiv}[1]{\href{https://arxiv.org/abs/#1}{\nolinkurl{https://arxiv.org/abs/#1}}}

\bibitem[{{Ananna} {et~al.}(2019){Ananna}, {Treister}, {Urry}, {Ricci},
  {Kirkpatrick}, {LaMassa}, {Buchner}, {Civano}, {Tremmel}, \&
  {Marchesi}}]{ananna2019}
{Ananna}, T.~T., {Treister}, E., {Urry}, C.~M., {et~al.} 2019, \apj, 871, 240,
  \dodoi{10.3847/1538-4357/aafb77}

\bibitem[{{Backhaus} {et~al.}(2022){Backhaus}, {Trump}, {Cleri}, {Simons},
  {Momcheva}, {Papovich}, {Estrada-Carpenter}, {Finkelstein}, {Matharu}, {Ji},
  {Weiner}, {Giavalisco}, \& {Jung}}]{backhaus2022}
{Backhaus}, B.~E., {Trump}, J.~R., {Cleri}, N.~J., {et~al.} 2022, \apj, 926,
  161, \dodoi{10.3847/1538-4357/ac3919}

\bibitem[{{Bell}(2003)}]{bell2003}
{Bell}, E.~F. 2003, \apj, 586, 794, \dodoi{10.1086/367829}

\bibitem[{{Bonato} {et~al.}(2017){Bonato}, {Sajina}, {De Zotti}, {McKinney},
  {Baronchelli}, {Negrello}, {Marchesini}, {Roebuck}, {Shipley}, {Kurinsky},
  {Pope}, {Noriega-Crespo}, {Yan}, \& {Kirkpatrick}}]{bonato2017}
{Bonato}, M., {Sajina}, A., {De Zotti}, G., {et~al.} 2017, \apj, 836, 171,
  \dodoi{10.3847/1538-4357/aa5c85}

\bibitem[{{Boquien} {et~al.}(2016){Boquien}, {Kennicutt}, {Calzetti}, {Dale},
  {Galametz}, {Sauvage}, {Croxall}, {Draine}, {Kirkpatrick}, {Kumari}, {Hunt},
  {De Looze}, {Pellegrini}, {Rela{\~n}o}, {Smith}, \&
  {Tabatabaei}}]{boquien2016}
{Boquien}, M., {Kennicutt}, R., {Calzetti}, D., {et~al.} 2016, \aap, 591, A6,
  \dodoi{10.1051/0004-6361/201527759}

\bibitem[{{Brammer} {et~al.}(2008){Brammer}, {van Dokkum}, \&
  {Coppi}}]{brammer2008}
{Brammer}, G.~B., {van Dokkum}, P.~G., \& {Coppi}, P. 2008, \apj, 686, 1503,
  \dodoi{10.1086/591786}

\bibitem[{{Calzetti} {et~al.}(2007){Calzetti}, {Kennicutt}, {Engelbracht},
  {Leitherer}, {Draine}, {Kewley}, {Moustakas}, {Sosey}, {Dale}, {Gordon},
  {Helou}, {Hollenbach}, {Armus}, {Bendo}, {Bot}, {Buckalew}, {Jarrett}, {Li},
  {Meyer}, {Murphy}, {Prescott}, {Regan}, {Rieke}, {Roussel}, {Sheth}, {Smith},
  {Thornley}, \& {Walter}}]{calzetti2007}
{Calzetti}, D., {Kennicutt}, R.~C., {Engelbracht}, C.~W., {et~al.} 2007, \apj,
  666, 870, \dodoi{10.1086/520082}

\bibitem[{{Caputi} {et~al.}(2007){Caputi}, {Lagache}, {Yan}, {Dole},
  {Bavouzet}, {Le Floc'h}, {Choi}, {Helou}, \& {Reddy}}]{caputi2007}
{Caputi}, K.~I., {Lagache}, G., {Yan}, L., {et~al.} 2007, \apj, 660, 97,
  \dodoi{10.1086/512667}

\bibitem[{{Chary} \& {Elbaz}(2001)}]{chary2001}
{Chary}, R., \& {Elbaz}, D. 2001, \apj, 556, 562, \dodoi{10.1086/321609}

\bibitem[{{Chary} {et~al.}(2004){Chary}, {Casertano}, {Dickinson}, {Ferguson},
  {Eisenhardt}, {Elbaz}, {Grogin}, {Moustakas}, {Reach}, \& {Yan}}]{chary2004}
{Chary}, R., {Casertano}, S., {Dickinson}, M.~E., {et~al.} 2004, \apjs, 154,
  80, \dodoi{10.1086/423307}

\bibitem[{{Cleri} {et~al.}(2023{\natexlab{a}}){Cleri}, {Yang}, {Papovich},
  {Trump}, {Backhaus}, {Estrada-Carpenter}, {Finkelstein}, {Giavalisco},
  {Hutchison}, {Ji}, {Jung}, {Matharu}, {Momcheva}, {Olivier}, {Simons}, \&
  {Weiner}}]{cleri2023a}
{Cleri}, N.~J., {Yang}, G., {Papovich}, C., {et~al.} 2023{\natexlab{a}}, \apj,
  948, 112, \dodoi{10.3847/1538-4357/acc1e6}

\bibitem[{{Cleri} {et~al.}(2023{\natexlab{b}}){Cleri}, {Olivier}, {Hutchison},
  {Papovich}, {Trump}, {Amorin}, {Backhaus}, {Berg}, {Fernandez},
  {Finkelstein}, {Fujimoto}, {Hirschmann}, {Kartaltepe}, {Kocevski}, {Simons},
  {Wilkins}, \& {Yung}}]{cleri2023b}
{Cleri}, N.~J., {Olivier}, G.~M., {Hutchison}, T.~A., {et~al.}
  2023{\natexlab{b}}, arXiv e-prints, arXiv:2301.07745,
  \dodoi{10.48550/arXiv.2301.07745}

\bibitem[{{Cortzen} {et~al.}(2019){Cortzen}, {Garrett}, {Magdis}, {Rigopoulou},
  {Valentino}, {Pereira-Santaella}, {Combes}, {Alonso-Herrero}, {Toft},
  {Daddi}, {Elbaz}, {G{\'o}mez-Guijarro}, {Stockmann}, {Huang}, \&
  {Kramer}}]{cortzen2019}
{Cortzen}, I., {Garrett}, J., {Magdis}, G., {et~al.} 2019, \mnras, 482, 1618,
  \dodoi{10.1093/mnras/sty2777}

\bibitem[{{Dahlen} {et~al.}(2013){Dahlen}, {Mobasher}, {Faber}, {Ferguson},
  {Barro}, {Finkelstein}, {Finlator}, {Fontana}, {Gruetzbauch}, {Johnson},
  {Pforr}, {Salvato}, {Wiklind}, {Wuyts}, {Acquaviva}, {Dickinson}, {Guo},
  {Huang}, {Huang}, {Newman}, {Bell}, {Conselice}, {Galametz}, {Gawiser},
  {Giavalisco}, {Grogin}, {Hathi}, {Kocevski}, {Koekemoer}, {Koo}, {Lee},
  {McGrath}, {Papovich}, {Peth}, {Ryan}, {Somerville}, {Weiner}, \&
  {Wilson}}]{dahlen2013}
{Dahlen}, T., {Mobasher}, B., {Faber}, S.~M., {et~al.} 2013, \apj, 775, 93,
  \dodoi{10.1088/0004-637X/775/2/93}

\bibitem[{{Dalcanton} {et~al.}(2004){Dalcanton}, {Yoachim}, \&
  {Bernstein}}]{dalcanton2004}
{Dalcanton}, J.~J., {Yoachim}, P., \& {Bernstein}, R.~A. 2004, \apj, 608, 189,
  \dodoi{10.1086/386358}

\bibitem[{{Dale} \& {Helou}(2002)}]{dale2002}
{Dale}, D.~A., \& {Helou}, G. 2002, \apj, 576, 159, \dodoi{10.1086/341632}

\bibitem[{{Del Moro} {et~al.}(2016){Del Moro}, {Alexander}, {Bauer}, {Daddi},
  {Kocevski}, {McIntosh}, {Stanley}, {Brandt}, {Elbaz}, {Harrison}, {Luo},
  {Mullaney}, \& {Xue}}]{delmoro2016}
{Del Moro}, A., {Alexander}, D.~M., {Bauer}, F.~E., {et~al.} 2016, \mnras, 456,
  2105, \dodoi{10.1093/mnras/stv2748}

\bibitem[{{Delvecchio} {et~al.}(2014){Delvecchio}, {Gruppioni}, {Pozzi},
  {Berta}, {Zamorani}, {Cimatti}, {Lutz}, {Scott}, {Vignali}, {Cresci},
  {Feltre}, {Cooray}, {Vaccari}, {Fritz}, {Le Floc'h}, {Magnelli}, {Popesso},
  {Oliver}, {Bock}, {Carollo}, {Contini}, {Le F{\'e}vre}, {Lilly}, {Mainieri},
  {Renzini}, \& {Scodeggio}}]{delvecchio2014}
{Delvecchio}, I., {Gruppioni}, C., {Pozzi}, F., {et~al.} 2014, \mnras, 439,
  2736, \dodoi{10.1093/mnras/stu130}

\bibitem[{{Donley} {et~al.}(2012){Donley}, {Koekemoer}, {Brusa}, {Capak},
  {Cardamone}, {Civano}, {Ilbert}, {Impey}, {Kartaltepe}, {Miyaji}, {Salvato},
  {Sanders}, {Trump}, \& {Zamorani}}]{donley2012}
{Donley}, J.~L., {Koekemoer}, A.~M., {Brusa}, M., {et~al.} 2012, \apj, 748,
  142, \dodoi{10.1088/0004-637X/748/2/142}

\bibitem[{{Draine} \& {Li}(2007)}]{draine2007}
{Draine}, B.~T., \& {Li}, A. 2007, \apj, 657, 810, \dodoi{10.1086/511055}

\bibitem[{{Elbaz} {et~al.}(2007){Elbaz}, {Daddi}, {Le Borgne}, {Dickinson},
  {Alexander}, {Chary}, {Starck}, {Brandt}, {Kitzbichler}, {MacDonald},
  {Nonino}, {Popesso}, {Stern}, \& {Vanzella}}]{elbaz2007}
{Elbaz}, D., {Daddi}, E., {Le Borgne}, D., {et~al.} 2007, \aap, 468, 33,
  \dodoi{10.1051/0004-6361:20077525}

\bibitem[{{Elbaz} {et~al.}(2011){Elbaz}, {Dickinson}, {Hwang},
  {D{\'\i}az-Santos}, {Magdis}, {Magnelli}, {Le Borgne}, {Galliano},
  {Pannella}, \& {Chanial}}]{elbaz2011}
{Elbaz}, D., {Dickinson}, M., {Hwang}, H.~S., {et~al.} 2011, \aap, 533, A119,
  \dodoi{10.1051/0004-6361/201117239}

\bibitem[{{Elitzur} {et~al.}(2014){Elitzur}, {Ho}, \& {Trump}}]{elitzur2014}
{Elitzur}, M., {Ho}, L.~C., \& {Trump}, J.~R. 2014, \mnras, 438, 3340,
  \dodoi{10.1093/mnras/stt2445}

\bibitem[{{Fazio} {et~al.}(2004){Fazio}, {Ashby}, {Barmby}, {Hora}, {Huang},
  {Pahre}, {Wang}, {Willner}, {Arendt}, {Moseley}, {Brodwin}, {Eisenhardt},
  {Stern}, {Tollestrup}, \& {Wright}}]{fazio2004}
{Fazio}, G.~G., {Ashby}, M.~L.~N., {Barmby}, P., {et~al.} 2004, \apjs, 154, 39,
  \dodoi{10.1086/422585}

\bibitem[{{Finkelstein} {et~al.}(2022){Finkelstein}, {Bagley}, {Song},
  {Larson}, {Papovich}, {Dickinson}, {Finkelstein}, {Koekemoer}, {Pirzkal},
  {Somerville}, {Yung}, {Behroozi}, {Ferguson}, {Giavalisco}, {Grogin},
  {Hathi}, {Hutchison}, {Jung}, {Kocevski}, {Kawinwanichakij}, {Rojas-Ruiz},
  {Ryan}, {Snyder}, \& {Tacchella}}]{finkelstein2022}
{Finkelstein}, S.~L., {Bagley}, M., {Song}, M., {et~al.} 2022, \apj, 928, 52,
  \dodoi{10.3847/1538-4357/ac3aed}

\bibitem[{{Gardner} {et~al.}(2006){Gardner}, {Mather}, {Clampin}, {Doyon},
  {Greenhouse}, {Hammel}, {Hutchings}, {Jakobsen}, {Lilly}, {Long}, {Lunine},
  {McCaughrean}, {Mountain}, {Nella}, {Rieke}, {Rieke}, {Rix}, {Smith},
  {Sonneborn}, {Stiavelli}, {Stockman}, {Windhorst}, \& {Wright}}]{gardner2006}
{Gardner}, J.~P., {Mather}, J.~C., {Clampin}, M., {et~al.} 2006, \ssr, 123,
  485, \dodoi{10.1007/s11214-006-8315-7}

\bibitem[{{Grogin} {et~al.}(2011){Grogin}, {Kocevski}, {Faber}, {Ferguson},
  {Koekemoer}, {Riess}, {Acquaviva}, {Alexander}, {Almaini}, {Ashby}, {Barden},
  {Bell}, {Bournaud}, {Brown}, {Caputi}, {Casertano}, {Cassata}, {Castellano},
  {Challis}, {Chary}, {Cheung}, {Cirasuolo}, {Conselice}, {Roshan Cooray},
  {Croton}, {Daddi}, {Dahlen}, {Dav{\'e}}, {de Mello}, {Dekel}, {Dickinson},
  {Dolch}, {Donley}, {Dunlop}, {Dutton}, {Elbaz}, {Fazio}, {Filippenko},
  {Finkelstein}, {Fontana}, {Gardner}, {Garnavich}, {Gawiser}, {Giavalisco},
  {Grazian}, {Guo}, {Hathi}, {H{\"a}ussler}, {Hopkins}, {Huang}, {Huang},
  {Jha}, {Kartaltepe}, {Kirshner}, {Koo}, {Lai}, {Lee}, {Li}, {Lotz}, {Lucas},
  {Madau}, {McCarthy}, {McGrath}, {McIntosh}, {McLure}, {Mobasher},
  {Moustakas}, {Mozena}, {Nandra}, {Newman}, {Niemi}, {Noeske}, {Papovich},
  {Pentericci}, {Pope}, {Primack}, {Rajan}, {Ravindranath}, {Reddy}, {Renzini},
  {Rix}, {Robaina}, {Rodney}, {Rosario}, {Rosati}, {Salimbeni}, {Scarlata},
  {Siana}, {Simard}, {Smidt}, {Somerville}, {Spinrad}, {Straughn}, {Strolger},
  {Telford}, {Teplitz}, {Trump}, {van der Wel}, {Villforth}, {Wechsler},
  {Weiner}, {Wiklind}, {Wild}, {Wilson}, {Wuyts}, {Yan}, \& {Yun}}]{grogin2011}
{Grogin}, N.~A., {Kocevski}, D.~D., {Faber}, S.~M., {et~al.} 2011, \apjs, 197,
  35, \dodoi{10.1088/0067-0049/197/2/35}

\bibitem[{{Hao} {et~al.}(2010){Hao}, {Elvis}, {Civano}, {Lanzuisi}, {Brusa},
  {Lusso}, {Zamorani}, {Comastri}, {Bongiorno}, {Impey}, {Koekemoer}, {Le
  Floc'h}, {Salvato}, {Sanders}, {Trump}, \& {Vignali}}]{hao2010}
{Hao}, H., {Elvis}, M., {Civano}, F., {et~al.} 2010, \apjl, 724, L59,
  \dodoi{10.1088/2041-8205/724/1/L59}

\bibitem[{{Hatcher} {et~al.}(2021){Hatcher}, {Kirkpatrick}, {Fornasini},
  {Civano}, {Lambrides}, {Kocesvski}, {Carroll}, {Giavalisco}, {Hickox}, \&
  {Ji}}]{hatcher2021}
{Hatcher}, C., {Kirkpatrick}, A., {Fornasini}, F., {et~al.} 2021, \aj, 162, 65,
  \dodoi{10.3847/1538-3881/ac0530}

\bibitem[{{Hickox} \& {Alexander}(2018)}]{hickox2018}
{Hickox}, R.~C., \& {Alexander}, D.~M. 2018, \araa, 56, 625,
  \dodoi{10.1146/annurev-astro-081817-051803}

\bibitem[{{Holwerda} {et~al.}(2009){Holwerda}, {Keel}, {Williams}, {Dalcanton},
  \& {de Jong}}]{holwerda2009}
{Holwerda}, B.~W., {Keel}, W.~C., {Williams}, B., {Dalcanton}, J.~J., \& {de
  Jong}, R.~S. 2009, \aj, 137, 3000, \dodoi{10.1088/0004-6256/137/2/3000}

\bibitem[{{Holwerda} {et~al.}(2012){Holwerda}, {Bianchi}, {B{\"o}ker},
  {Radburn-Smith}, {de Jong}, {Baes}, {van der Kruit}, {Xilouris}, {Gordon}, \&
  {Dalcanton}}]{holwerda2012}
{Holwerda}, B.~W., {Bianchi}, S., {B{\"o}ker}, T., {et~al.} 2012, \aap, 541,
  L5, \dodoi{10.1051/0004-6361/201118615}

\bibitem[{{Holwerda} {et~al.}(2021){Holwerda}, {Wu}, {Keel}, {Young},
  {Mullins}, {Hinz}, {Ford}, {Barmby}, {Chandar}, {Bailin}, {Peek},
  {Pickering}, \& {B{\"o}ker}}]{holwerda2021}
{Holwerda}, B.~W., {Wu}, J.~F., {Keel}, W.~C., {et~al.} 2021, \apj, 914, 142,
  \dodoi{10.3847/1538-4357/abffcc}

\bibitem[{{Jarrett} {et~al.}(2011){Jarrett}, {Cohen}, {Masci}, {Wright},
  {Stern}, {Benford}, {Blain}, {Carey}, {Cutri}, {Eisenhardt}, {Lonsdale},
  {Mainzer}, {Marsh}, {Padgett}, {Petty}, {Ressler}, {Skrutskie}, {Stanford},
  {Surace}, {Tsai}, {Wheelock}, \& {Yan}}]{jarrett2011}
{Jarrett}, T.~H., {Cohen}, M., {Masci}, F., {et~al.} 2011, \apj, 735, 112,
  \dodoi{10.1088/0004-637X/735/2/112}

\bibitem[{{Jin} {et~al.}(2018){Jin}, {Daddi}, {Liu}, {Smol{\v{c}}i{\'c}},
  {Schinnerer}, {Calabr{\`o}}, {Gu}, {Delhaize}, {Delvecchio}, {Gao},
  {Salvato}, {Puglisi}, {Dickinson}, {Bertoldi}, {Sargent}, {Novak}, {Magdis},
  {Aretxaga}, {Wilson}, \& {Capak}}]{jin2018}
{Jin}, S., {Daddi}, E., {Liu}, D., {et~al.} 2018, \apj, 864, 56,
  \dodoi{10.3847/1538-4357/aad4af}

\bibitem[{{Kartaltepe} {et~al.}(2012){Kartaltepe}, {Dickinson}, {Alexander},
  {Bell}, {Dahlen}, {Elbaz}, {Faber}, {Lotz}, {McIntosh}, {Wiklind}, {Altieri},
  {Aussel}, {Bethermin}, {Bournaud}, {Charmandaris}, {Conselice}, {Cooray},
  {Dannerbauer}, {Dav{\'e}}, {Dunlop}, {Dekel}, {Ferguson}, {Grogin}, {Hwang},
  {Ivison}, {Kocevski}, {Koekemoer}, {Koo}, {Lai}, {Leiton}, {Lucas}, {Lutz},
  {Magdis}, {Magnelli}, {Morrison}, {Mozena}, {Mullaney}, {Newman}, {Pope},
  {Popesso}, {van der Wel}, {Weiner}, \& {Wuyts}}]{kartaltepe2012}
{Kartaltepe}, J.~S., {Dickinson}, M., {Alexander}, D.~M., {et~al.} 2012, \apj,
  757, 23, \dodoi{10.1088/0004-637X/757/1/23}

\bibitem[{{Kennicutt} \& {Evans}(2012)}]{kennicutt2012}
{Kennicutt}, R.~C., \& {Evans}, N.~J. 2012, \araa, 50, 531,
  \dodoi{10.1146/annurev-astro-081811-125610}

\bibitem[{{Kennicutt}(1998)}]{kennicutt1998}
{Kennicutt}, Jr., R.~C. 1998, \apj, 498, 541, \dodoi{10.1086/305588}

\bibitem[{{Kirkpatrick} {et~al.}(2015){Kirkpatrick}, {Pope}, {Sajina},
  {Roebuck}, {Yan}, {Armus}, {D{\'{\i}}az-Santos}, \&
  {Stierwalt}}]{kirkpatrick2015}
{Kirkpatrick}, A., {Pope}, A., {Sajina}, A., {et~al.} 2015, \apj, 814, 9,
  \dodoi{10.1088/0004-637X/814/1/9}

\bibitem[{{Kirkpatrick} {et~al.}(2012){Kirkpatrick}, {Pope}, {Alexander},
  {Charmandaris}, {Daddi}, {Dickinson}, {Elbaz}, {Gabor}, {Hwang}, {Ivison},
  {Mullaney}, {Pannella}, {Scott}, {Altieri}, {Aussel}, {Bournaud}, {Buat},
  {Coia}, {Dannerbauer}, {Dasyra}, {Kartaltepe}, {Leiton}, {Lin}, {Magdis},
  {Magnelli}, {Morrison}, {Popesso}, \& {Valtchanov}}]{kirkpatrick2012}
{Kirkpatrick}, A., {Pope}, A., {Alexander}, D.~M., {et~al.} 2012, \apj, 759,
  139, \dodoi{10.1088/0004-637X/759/2/139}

\bibitem[{{Kirkpatrick} {et~al.}(2013){Kirkpatrick}, {Pope}, {Charmandaris},
  {Daddi}, {Elbaz}, {Hwang}, {Pannella}, {Scott}, {Altieri}, \&
  {Aussel}}]{kirkpatrick2013}
{Kirkpatrick}, A., {Pope}, A., {Charmandaris}, V., {et~al.} 2013, \apj, 763,
  123, \dodoi{10.1088/0004-637X/763/2/123}

\bibitem[{{Kirkpatrick} {et~al.}(2014){Kirkpatrick}, {Pope}, {Aretxaga},
  {Armus}, {Calzetti}, {Helou}, {Monta{\~n}a}, {Narayanan}, {Schloerb}, {Shi},
  {Vega}, \& {Yun}}]{kirkpatrick2014}
{Kirkpatrick}, A., {Pope}, A., {Aretxaga}, I., {et~al.} 2014, \apj, 796, 135,
  \dodoi{10.1088/0004-637X/796/2/135}

\bibitem[{{Kirkpatrick} {et~al.}(2017{\natexlab{a}}){Kirkpatrick}, {Alberts},
  {Pope}, {Barro}, {Bonato}, {Kocevski}, {P{\'e}rez-Gonz{\'a}lez}, {Rieke},
  {Rodr{\'{\i}}guez-Mu{\~n}oz}, {Sajina}, {Grogin}, {Mantha}, {Pandya},
  {Pforr}, {Salvato}, \& {Santini}}]{kirkpatrick2017}
{Kirkpatrick}, A., {Alberts}, S., {Pope}, A., {et~al.} 2017{\natexlab{a}},
  \apj, 849, 111, \dodoi{10.3847/1538-4357/aa911d}

\bibitem[{{Kirkpatrick} {et~al.}(2017{\natexlab{b}}){Kirkpatrick}, {Pope},
  {Sajina}, {Dale}, {D{\'\i}az-Santos}, {Hayward}, {Shi}, {Somerville},
  {Stierwalt}, {Armus}, {Kartaltepe}, {Kocevski}, {McIntosh}, {Sanders}, \&
  {Yan}}]{kirkpatrick2017b}
{Kirkpatrick}, A., {Pope}, A., {Sajina}, A., {et~al.} 2017{\natexlab{b}}, \apj,
  843, 71, \dodoi{10.3847/1538-4357/aa76dc}

\bibitem[{{Koekemoer} {et~al.}(2011){Koekemoer}, {Faber}, {Ferguson}, {Grogin},
  {Kocevski}, {Koo}, {Lai}, {Lotz}, {Lucas}, {McGrath}, {Ogaz}, {Rajan},
  {Riess}, {Rodney}, {Strolger}, {Casertano}, {Castellano}, {Dahlen},
  {Dickinson}, {Dolch}, {Fontana}, {Giavalisco}, {Grazian}, {Guo}, {Hathi},
  {Huang}, {van der Wel}, {Yan}, {Acquaviva}, {Alexander}, {Almaini}, {Ashby},
  {Barden}, {Bell}, {Bournaud}, {Brown}, {Caputi}, {Cassata}, {Challis},
  {Chary}, {Cheung}, {Cirasuolo}, {Conselice}, {Roshan Cooray}, {Croton},
  {Daddi}, {Dav{\'e}}, {de Mello}, {de Ravel}, {Dekel}, {Donley}, {Dunlop},
  {Dutton}, {Elbaz}, {Fazio}, {Filippenko}, {Finkelstein}, {Frazer}, {Gardner},
  {Garnavich}, {Gawiser}, {Gruetzbauch}, {Hartley}, {H{\"a}ussler},
  {Herrington}, {Hopkins}, {Huang}, {Jha}, {Johnson}, {Kartaltepe},
  {Khostovan}, {Kirshner}, {Lani}, {Lee}, {Li}, {Madau}, {McCarthy},
  {McIntosh}, {McLure}, {McPartland}, {Mobasher}, {Moreira}, {Mortlock},
  {Moustakas}, {Mozena}, {Nandra}, {Newman}, {Nielsen}, {Niemi}, {Noeske},
  {Papovich}, {Pentericci}, {Pope}, {Primack}, {Ravindranath}, {Reddy},
  {Renzini}, {Rix}, {Robaina}, {Rosario}, {Rosati}, {Salimbeni}, {Scarlata},
  {Siana}, {Simard}, {Smidt}, {Snyder}, {Somerville}, {Spinrad}, {Straughn},
  {Telford}, {Teplitz}, {Trump}, {Vargas}, {Villforth}, {Wagner}, {Wandro},
  {Wechsler}, {Weiner}, {Wiklind}, {Wild}, {Wilson}, {Wuyts}, \&
  {Yun}}]{koekemoer2011}
{Koekemoer}, A.~M., {Faber}, S.~M., {Ferguson}, H.~C., {et~al.} 2011, \apjs,
  197, 36, \dodoi{10.1088/0067-0049/197/2/36}

\bibitem[{{Kroupa}(2001)}]{kroupa2001}
{Kroupa}, P. 2001, \mnras, 322, 231, \dodoi{10.1046/j.1365-8711.2001.04022.x}

\bibitem[{{Lacy} {et~al.}(2004){Lacy}, {Storrie-Lombardi}, {Sajina},
  {Appleton}, {Armus}, {Chapman}, {Choi}, {Fadda}, {Fang}, {Frayer},
  {Heinrichsen}, {Helou}, {Im}, {Marleau}, {Masci}, {Shupe}, {Soifer},
  {Surace}, {Teplitz}, {Wilson}, \& {Yan}}]{lacy2004}
{Lacy}, M., {Storrie-Lombardi}, L.~J., {Sajina}, A., {et~al.} 2004, \apjs, 154,
  166, \dodoi{10.1086/422816}

\bibitem[{{Langeroodi} \& {Hjorth}(2023)}]{langeroodi2023}
{Langeroodi}, D., \& {Hjorth}, J. 2023, \apjl, 946, L40,
  \dodoi{10.3847/2041-8213/acc1e0}

\bibitem[{{Leja} {et~al.}(2019){Leja}, {Carnall}, {Johnson}, {Conroy}, \&
  {Speagle}}]{leja2019}
{Leja}, J., {Carnall}, A.~C., {Johnson}, B.~D., {Conroy}, C., \& {Speagle},
  J.~S. 2019, \apj, 876, 3, \dodoi{10.3847/1538-4357/ab133c}

\bibitem[{{Liu} {et~al.}(2018){Liu}, {Daddi}, {Dickinson}, {Owen}, {Pannella},
  {Sargent}, {B{\'e}thermin}, {Magdis}, {Gao}, {Shu}, {Wang}, {Jin}, \&
  {Inami}}]{liu2018}
{Liu}, D., {Daddi}, E., {Dickinson}, M., {et~al.} 2018, \apj, 853, 172,
  \dodoi{10.3847/1538-4357/aaa600}

\bibitem[{{Lyu} \& {Rieke}(2022)}]{lyu2022}
{Lyu}, J., \& {Rieke}, G. 2022, Universe, 8, 304,
  \dodoi{10.3390/universe8060304}

\bibitem[{{Lyu} {et~al.}(2017){Lyu}, {Rieke}, \& {Shi}}]{lyu2017b}
{Lyu}, J., {Rieke}, G.~H., \& {Shi}, Y. 2017, \apj, 835, 257,
  \dodoi{10.3847/1538-4357/835/2/257}

\bibitem[{{Magnelli} {et~al.}(2009){Magnelli}, {Elbaz}, {Chary}, {Dickinson},
  {Le Borgne}, {Frayer}, \& {Willmer}}]{magnelli2009}
{Magnelli}, B., {Elbaz}, D., {Chary}, R.~R., {et~al.} 2009, \aap, 496, 57,
  \dodoi{10.1051/0004-6361:200811443}

\bibitem[{{Magnelli} {et~al.}(2011){Magnelli}, {Elbaz}, {Chary}, {Dickinson},
  {Le Borgne}, {Frayer}, \& {Willmer}}]{magnelli2011}
---. 2011, \aap, 528, A35, \dodoi{10.1051/0004-6361/200913941}

\bibitem[{{Magnelli} {et~al.}(2023){Magnelli}, {G{\'o}mez-Guijarro}, {Elbaz},
  {Daddi}, {Papovich}, {Shen}, {Arrabal Haro}, {Bagley}, {Bell}, {Buat},
  {Costantin}, {Dickinson}, {Finkelstein}, {Gardner}, {Jim{\'e}nez-Andrade},
  {Kartaltepe}, {Koekemoer}, {Lyu}, {P{\'e}rez-Gonz{\'a}lez}, {Pirzkal},
  {Tacchella}, {de la Vega}, {Wuyts}, {Yang}, {Yung}, \&
  {Zavala}}]{magnelli2023}
{Magnelli}, B., {G{\'o}mez-Guijarro}, C., {Elbaz}, D., {et~al.} 2023, arXiv
  e-prints, arXiv:2305.19331, \dodoi{10.48550/arXiv.2305.19331}

\bibitem[{{Mason} {et~al.}(2012){Mason}, {Lopez-Rodriguez}, {Packham},
  {Alonso-Herrero}, {Levenson}, {Radomski}, {Ramos Almeida}, {Colina},
  {Elitzur}, {Aretxaga}, {Roche}, \& {Oi}}]{mason2012}
{Mason}, R.~E., {Lopez-Rodriguez}, E., {Packham}, C., {et~al.} 2012, \aj, 144,
  11, \dodoi{10.1088/0004-6256/144/1/11}

\bibitem[{{McKinney} {et~al.}(2020){McKinney}, {Pope}, {Armus}, {Chary},
  {D{\'\i}az-Santos}, {Dickinson}, \& {Kirkpatrick}}]{mckinney2020}
{McKinney}, J., {Pope}, A., {Armus}, L., {et~al.} 2020, \apj, 892, 119,
  \dodoi{10.3847/1538-4357/ab77b9}

\bibitem[{{Merlin} {et~al.}(2015){Merlin}, {Fontana}, {Ferguson}, {Dunlop},
  {Elbaz}, {Bourne}, {Bruce}, {Buitrago}, {Castellano}, {Schreiber}, {Grazian},
  {McLure}, {Okumura}, {Shu}, {Wang}, {Amor{\'\i}n}, {Boutsia}, {Cappelluti},
  {Comastri}, {Derriere}, {Faber}, \& {Santini}}]{merlin2015}
{Merlin}, E., {Fontana}, A., {Ferguson}, H.~C., {et~al.} 2015, \aap, 582, A15,
  \dodoi{10.1051/0004-6361/201526471}

\bibitem[{{Messias} {et~al.}(2012){Messias}, {Afonso}, {Salvato}, {Mobasher},
  \& {Hopkins}}]{messias2012}
{Messias}, H., {Afonso}, J., {Salvato}, M., {Mobasher}, B., \& {Hopkins}, A.~M.
  2012, \apj, 754, 120, \dodoi{10.1088/0004-637X/754/2/120}

\bibitem[{{Murphy} {et~al.}(2011){Murphy}, {Condon}, {Schinnerer}, {Kennicutt},
  {Calzetti}, {Armus}, {Helou}, {Turner}, {Aniano}, {Beir{\~a}o}, {Bolatto},
  {Brandl}, {Croxall}, {Dale}, {Donovan Meyer}, {Draine}, {Engelbracht},
  {Hunt}, {Hao}, {Koda}, {Roussel}, {Skibba}, \& {Smith}}]{murphy2011}
{Murphy}, E.~J., {Condon}, J.~J., {Schinnerer}, E., {et~al.} 2011, \apj, 737,
  67, \dodoi{10.1088/0004-637X/737/2/67}

\bibitem[{{Negus} {et~al.}(2023){Negus}, {Comerford}, {S{\'a}nchez},
  {Revalski}, {Riffel}, {Bundy}, {Nevin}, \& {Rembold}}]{negus2023}
{Negus}, J., {Comerford}, J.~M., {S{\'a}nchez}, F.~M., {et~al.} 2023, \apj,
  945, 127, \dodoi{10.3847/1538-4357/acb772}

\bibitem[{{Newman} \& {Gruen}(2022)}]{newman2022}
{Newman}, J.~A., \& {Gruen}, D. 2022, \araa, 60, 363,
  \dodoi{10.1146/annurev-astro-032122-014611}

\bibitem[{{Noeske} {et~al.}(2007){Noeske}, {Weiner}, {Faber}, {Papovich},
  {Koo}, {Somerville}, {Bundy}, {Conselice}, {Newman}, {Schiminovich}, {Le
  Floc'h}, {Coil}, {Rieke}, {Lotz}, {Primack}, {Barmby}, {Cooper}, {Davis},
  {Ellis}, {Fazio}, {Guhathakurta}, {Huang}, {Kassin}, {Martin}, {Phillips},
  {Rich}, {Small}, {Willmer}, \& {Wilson}}]{noeske2007}
{Noeske}, K.~G., {Weiner}, B.~J., {Faber}, S.~M., {et~al.} 2007, \apjl, 660,
  L43, \dodoi{10.1086/517926}

\bibitem[{{Padovani} {et~al.}(2017){Padovani}, {Alexander}, {Assef}, {De
  Marco}, {Giommi}, {Hickox}, {Richards}, {Smol{\v{c}}i{\'c}},
  {Hatziminaoglou}, {Mainieri}, \& {Salvato}}]{padovani2017}
{Padovani}, P., {Alexander}, D.~M., {Assef}, R.~J., {et~al.} 2017, \aapr, 25,
  2, \dodoi{10.1007/s00159-017-0102-9}

\bibitem[{{Papovich} {et~al.}(2004){Papovich}, {Dole}, {Egami}, {Le Floc'h},
  {P{\'e}rez-Gonz{\'a}lez}, {Alonso-Herrero}, {Bai}, {Beichman}, {Blaylock},
  {Engelbracht}, {Gordon}, {Hines}, {Misselt}, {Morrison}, {Mould},
  {Muzerolle}, {Neugebauer}, {Richards}, {Rieke}, {Rieke}, {Rigby}, {Su}, \&
  {Young}}]{papovich2004}
{Papovich}, C., {Dole}, H., {Egami}, E., {et~al.} 2004, \apjs, 154, 70,
  \dodoi{10.1086/422880}

\bibitem[{{Papovich} {et~al.}(2022){Papovich}, {Cole}, {Yang}, {Finkelstein},
  {Barro}, {Buat}, {Burgarella}, {P{\'e}rez-Gonz{\'a}lez}, {Santini},
  {Seill{\'e}}, {Shen}, {Arrabal Haro}, {Bagley}, {Bell}, {Bisigello},
  {Calabr{\`o}}, {Casey}, {Castellano}, {Chworowsky}, {Cleri}, {Cooper},
  {Costantin}, {Dickinson}, {Ferguson}, {Fontana}, {Giavalisco}, {Grazian},
  {Grogin}, {Hathi}, {Holwerda}, {Hutchison}, {Kartaltepe}, {Kewley},
  {Kirkpatrick}, {Kocevski}, {Koekemoer}, {Larson}, {Long}, {Lucas},
  {Pentericci}, {Pirzkal}, {Ravindranath}, {Somerville}, {Trump}, {Urbano
  Stawinski}, {Weiner}, {Wilkins}, {Yung}, \& {Zavala}}]{papovich2023}
{Papovich}, C., {Cole}, J., {Yang}, G., {et~al.} 2022, arXiv e-prints,
  arXiv:2301.00027, \dodoi{10.48550/arXiv.2301.00027}

\bibitem[{{Peca} {et~al.}(2023){Peca}, {Cappelluti}, {Urry}, {LaMassa},
  {Marchesi}, {Ananna}, {Balokovi{\'c}}, {Sanders}, {Auge}, {Treister},
  {Powell}, {Turner}, {Kirkpatrick}, \& {Tian}}]{peca2023}
{Peca}, A., {Cappelluti}, N., {Urry}, C.~M., {et~al.} 2023, \apj, 943, 162,
  \dodoi{10.3847/1538-4357/acac28}

\bibitem[{{P{\'e}rez-Gonz{\'a}lez} {et~al.}(2005){P{\'e}rez-Gonz{\'a}lez},
  {Rieke}, {Egami}, {Alonso-Herrero}, {Dole}, {Papovich}, {Blaylock}, {Jones},
  {Rieke}, {Rigby}, {Barmby}, {Fazio}, {Huang}, \& {Martin}}]{perez2005}
{P{\'e}rez-Gonz{\'a}lez}, P.~G., {Rieke}, G.~H., {Egami}, E., {et~al.} 2005,
  \apj, 630, 82, \dodoi{10.1086/431894}

\bibitem[{{Petric} {et~al.}(2011){Petric}, {Armus}, {Howell}, {Chan},
  {Mazzarella}, {Evans}, {Surace}, {Sand ers}, {Appleton}, {Charmandaris},
  {D{\'\i}az-Santos}, {Frayer}, {Haan}, {Inami}, {Iwasawa}, {Kim}, {Madore},
  {Marshall}, {Spoon}, {Stierwalt}, {Sturm}, {U}, {Vavilkin}, \&
  {Veilleux}}]{petric2011}
{Petric}, A.~O., {Armus}, L., {Howell}, J., {et~al.} 2011, \apj, 730, 28,
  \dodoi{10.1088/0004-637X/730/1/28}

\bibitem[{{Pope} {et~al.}(2008){Pope}, {Chary}, {Alexander}, {Armus},
  {Dickinson}, {Elbaz}, {Frayer}, {Scott}, \& {Teplitz}}]{pope2008}
{Pope}, A., {Chary}, R.-R., {Alexander}, D.~M., {et~al.} 2008, \apj, 675, 1171,
  \dodoi{10.1086/527030}

\bibitem[{{Pope} {et~al.}(2013){Pope}, {Wagg}, {Frayer}, {Armus}, {Chary},
  {Daddi}, {Desai}, {Dickinson}, {Elbaz}, {Gabor}, \& {Kirkpatrick}}]{pope2013}
{Pope}, A., {Wagg}, J., {Frayer}, D., {et~al.} 2013, \apj, 772, 92,
  \dodoi{10.1088/0004-637X/772/2/92}

\bibitem[{{Rieke} {et~al.}(2009){Rieke}, {Alonso-Herrero}, {Weiner},
  {P{\'e}rez-Gonz{\'a}lez}, {Blaylock}, {Donley}, \& {Marcillac}}]{rieke2009}
{Rieke}, G.~H., {Alonso-Herrero}, A., {Weiner}, B.~J., {et~al.} 2009, \apj,
  692, 556, \dodoi{10.1088/0004-637X/692/1/556}

\bibitem[{{Rieke} {et~al.}(2015){Rieke}, {Wright}, {B{\"o}ker}, {Bouwman},
  {Colina}, {Glasse}, {Gordon}, {Greene}, {G{\"u}del}, {Henning}, {Justtanont},
  {Lagage}, {Meixner}, {N{\o}rgaard-Nielsen}, {Ray}, {Ressler}, {van Dishoeck},
  \& {Waelkens}}]{rieke2015}
{Rieke}, G.~H., {Wright}, G.~S., {B{\"o}ker}, T., {et~al.} 2015, \pasp, 127,
  584, \dodoi{10.1086/682252}

\bibitem[{{Rodighiero} {et~al.}(2010){Rodighiero}, {Vaccari}, {Franceschini},
  {Tresse}, {Le Fevre}, {Le Brun}, {Mancini}, {Matute}, {Cimatti}, {Marchetti},
  {Ilbert}, {Arnouts}, {Bolzonella}, {Zucca}, {Bardelli}, {Lonsdale}, {Shupe},
  {Surace}, {Rowan-Robinson}, {Garilli}, {Zamorani}, {Pozzetti}, {Bondi}, {de
  la Torre}, {Vergani}, {Santini}, {Grazian}, \& {Fontana}}]{rodighiero2010}
{Rodighiero}, G., {Vaccari}, M., {Franceschini}, A., {et~al.} 2010, \aap, 515,
  A8, \dodoi{10.1051/0004-6361/200912058}

\bibitem[{{Sajina} {et~al.}(2012){Sajina}, {Yan}, {Fadda}, {Dasyra}, \&
  {Huynh}}]{sajina2012}
{Sajina}, A., {Yan}, L., {Fadda}, D., {Dasyra}, K., \& {Huynh}, M. 2012, \apj,
  757, 13, \dodoi{10.1088/0004-637X/757/1/13}

\bibitem[{{Santini} {et~al.}(2015){Santini}, {Ferguson}, {Fontana}, {Mobasher},
  {Barro}, {Castellano}, {Finkelstein}, {Grazian}, {Hsu}, {Lee}, {Lee},
  {Pforr}, {Salvato}, {Wiklind}, {Wuyts}, {Almaini}, {Cooper}, {Galametz},
  {Weiner}, {Amorin}, {Boutsia}, {Conselice}, {Dahlen}, {Dickinson},
  {Giavalisco}, {Grogin}, {Guo}, {Hathi}, {Kocevski}, {Koekemoer},
  {Kurczynski}, {Merlin}, {Mortlock}, {Newman}, {Paris}, {Pentericci},
  {Simons}, \& {Willner}}]{santini2015}
{Santini}, P., {Ferguson}, H.~C., {Fontana}, A., {et~al.} 2015, \apj, 801, 97,
  \dodoi{10.1088/0004-637X/801/2/97}

\bibitem[{{Schreiber} {et~al.}(2018){Schreiber}, {Elbaz}, {Pannella}, {Ciesla},
  {Wang}, \& {Franco}}]{schreiber2018}
{Schreiber}, C., {Elbaz}, D., {Pannella}, M., {et~al.} 2018, \aap, 609, A30,
  \dodoi{10.1051/0004-6361/201731506}

\bibitem[{{Shen} {et~al.}(2023){Shen}, {Papovich}, {Yang}, {Matharu}, {Wang},
  {Magnelli}, {Elbaz}, {Jogee}, {Alavi}, {Arrabal Haro}, {Backhaus}, {Bagley},
  {Bell}, {Bisigello}, {Calabr{\`o}}, {Cooper}, {Costantin}, {Daddi},
  {Dickinson}, {Finkelstein}, {Fujimoto}, {Giavalisco}, {Grogin}, {Guo},
  {Holwerda}, {Kartaltepe}, {Koekemoer}, {Kurczynski}, {Lucas},
  {Pe{\'r}ez-Gonza{\'l}ez}, {Pirzkal}, {Prichard}, {Rafelski}, {Ronayne},
  {Simons}, {Sunnquist}, {Teplitz}, {Trump}, {Weiner}, {Windhorst}, \&
  {Yung}}]{shen2023}
{Shen}, L., {Papovich}, C., {Yang}, G., {et~al.} 2023, arXiv e-prints,
  arXiv:2301.05727, \dodoi{10.48550/arXiv.2301.05727}

\bibitem[{{Stefanon} {et~al.}(2017){Stefanon}, {Yan}, {Mobasher}, {Barro},
  {Donley}, {Fontana}, {Hemmati}, {Koekemoer}, {Lee}, {Lee}, {Nayyeri}, {Peth},
  {Pforr}, {Salvato}, {Wiklind}, {Wuyts}, {Ashby}, {Castellano}, {Conselice},
  {Cooper}, {Cooray}, {Dolch}, {Ferguson}, {Galametz}, {Giavalisco}, {Guo},
  {Willner}, {Dickinson}, {Faber}, {Fazio}, {Gardner}, {Gawiser}, {Grazian},
  {Grogin}, {Kocevski}, {Koo}, {Lee}, {Lucas}, {McGrath}, {Nandra}, {Newman},
  \& {van der Wel}}]{stefanon2017}
{Stefanon}, M., {Yan}, H., {Mobasher}, B., {et~al.} 2017, \apjs, 229, 32,
  \dodoi{10.3847/1538-4365/aa66cb}

\bibitem[{{Stern} {et~al.}(2005){Stern}, {Eisenhardt}, {Gorjian}, {Kochanek},
  {Caldwell}, {Eisenstein}, {Brodwin}, {Brown}, {Cool}, {Dey}, {Green},
  {Jannuzi}, {Murray}, {Pahre}, \& {Willner}}]{stern2005}
{Stern}, D., {Eisenhardt}, P., {Gorjian}, V., {et~al.} 2005, \apj, 631, 163,
  \dodoi{10.1086/432523}

\bibitem[{{Stone} {et~al.}(2022){Stone}, {Pope}, {McKinney}, {Armus},
  {D{\'\i}az-Santos}, {Inami}, {Kirkpatrick}, \& {Stierwalt}}]{stone2022}
{Stone}, M., {Pope}, A., {McKinney}, J., {et~al.} 2022, \apj, 934, 27,
  \dodoi{10.3847/1538-4357/ac778b}

\bibitem[{{Tardugno Poleo} {et~al.}(2023){Tardugno Poleo}, {Finkelstein},
  {Leung}, {Mentuch Cooper}, {Gebhardt}, {Farrow}, {Gawiser}, {Zeimann},
  {Schneider}, {Morabito}, {Mock}, \& {Liu}}]{poleo2023}
{Tardugno Poleo}, V., {Finkelstein}, S.~L., {Leung}, G., {et~al.} 2023, \aj,
  165, 153, \dodoi{10.3847/1538-3881/acba92}

\bibitem[{{Trewhella} {et~al.}(2000){Trewhella}, {Davies}, {Alton}, {Bianchi},
  \& {Madore}}]{trewhella2000}
{Trewhella}, M., {Davies}, J.~I., {Alton}, P.~B., {Bianchi}, S., \& {Madore},
  B.~F. 2000, \apj, 543, 153, \dodoi{10.1086/317083}

\bibitem[{{Volonteri} {et~al.}(2017){Volonteri}, {Reines}, {Atek}, {Stark}, \&
  {Trebitsch}}]{volonteri2017}
{Volonteri}, M., {Reines}, A.~E., {Atek}, H., {Stark}, D.~P., \& {Trebitsch},
  M. 2017, \apj, 849, 155, \dodoi{10.3847/1538-4357/aa93f1}

\bibitem[{{Xilouris} {et~al.}(2004){Xilouris}, {Madden}, {Galliano}, {Vigroux},
  \& {Sauvage}}]{xilouris2004}
{Xilouris}, E.~M., {Madden}, S.~C., {Galliano}, F., {Vigroux}, L., \&
  {Sauvage}, M. 2004, \aap, 416, 41, \dodoi{10.1051/0004-6361:20034020}

\bibitem[{{Yang} {et~al.}(2021){Yang}, {Papovich}, {Bagley}, {Buat},
  {Burgarella}, {Dickinson}, {Elbaz}, {Finkelstein}, {Fontana}, {Grogin},
  {Jung}, {Kartaltepe}, {Kirkpatrick}, {Koekemoer}, {P{\'e}rez-Gonz{\'a}lez},
  {Pirzkal}, \& {Yung}}]{yang2021}
{Yang}, G., {Papovich}, C., {Bagley}, M.~B., {et~al.} 2021, \apj, 908, 144,
  \dodoi{10.3847/1538-4357/abd6c1}

\bibitem[{{Yang} {et~al.}(2023{\natexlab{a}}){Yang}, {Caputi}, {Papovich},
  {Arrabal Haro}, {Bagley}, {Behroozi}, {Bell}, {Bisigello}, {Buat},
  {Burgarella}, {Cheng}, {Cleri}, {Dave}, {Dickinson}, {Elbaz}, {Ferguson},
  {Finkelstein}, {Grogin}, {Hathi}, {Hirschmann}, {Holwerda},
  {Huertas-Company}, {Hutchison}, {Iani}, {Kartaltepe}, {Kirkpatrick},
  {Kocevski}, {Koekemoer}, {Kokorev}, {Larson}, {Lucas}, {Perez-Gonzalez},
  {Rinaldi}, {Shen}, {Trump}, {de la Vega}, {Yung}, \& {Zavala}}]{yang2023}
{Yang}, G., {Caputi}, K.~I., {Papovich}, C., {et~al.} 2023{\natexlab{a}}, arXiv
  e-prints, arXiv:2303.11736, \dodoi{10.48550/arXiv.2303.11736}

\bibitem[{{Yang} {et~al.}(2023{\natexlab{b}}){Yang}, {Papovich}, {Bagley},
  {Ferguson}, {Finkelstein}, {Koekemoer}, {P{\'e}rez-Gonz{\'a}lez}, {Arrabal
  Haro}, {Bisigello}, {Caputi}, {Cheng}, {Costantin}, {Dickinson}, {Fontana},
  {Gardner}, {Grazian}, {Grogin}, {Harish}, {Holwerda}, {Iani}, {Kartaltepe},
  {Kewley}, {Kirkpatrick}, {Kocevski}, {Kokorev}, {Lotz}, {Lucas},
  {Navarro-Carrera}, {Pentericci}, {Pirzkal}, {Ravindranath}, {Rinaldi},
  {Shen}, {Somerville}, {Trump}, {de la Vega}, {Wilkins}, \&
  {Yung}}]{yang2023b}
{Yang}, G., {Papovich}, C., {Bagley}, M., {et~al.} 2023{\natexlab{b}}, arXiv
  e-prints, arXiv:2307.14509, \dodoi{10.48550/arXiv.2307.14509}

\bibitem[{{Yuan} \& {Narayan}(2014)}]{yuan2014}
{Yuan}, F., \& {Narayan}, R. 2014, \araa, 52, 529,
  \dodoi{10.1146/annurev-astro-082812-141003}

\end{thebibliography}

\end{document}